\documentclass[twocolumn]{aastex631}
\usepackage{amsmath}
\newcommand{\e}[2]{$\rm #1 \times 10^{#2}$}

\newcommand{\msun}{M$_{\sun}$}
\newcommand{\rsun}{R$_{\sun}$}

\received{April 4, 2023}
\revised{May 25, 2023}
\accepted{\today}

\submitjournal{APJ}

\shorttitle{Radiation Transport for Interacting Transients with SuperLite}
\shortauthors{Wagle et al.}
\graphicspath{{./}{figures/}}

\begin{document}

\title{Monte Carlo Radiation Transport for Astrophysical Transients Powered by Circumstellar Interaction}

\correspondingauthor{Gururaj Wagle}
\email{gwagle1@lsu.edu}

\author[0000-0002-3356-5855]{Gururaj Wagle}
\affiliation{Department of Physics and Astronomy, Louisiana State University, Baton Rouge, LA 70803, USA}

\author[0000-0002-8179-1654]{Emmanouil Chatzopoulos}
\affiliation{Department of Physics and Astronomy, Louisiana State University, Baton Rouge, LA 70803, USA}
\affiliation{Institute of Astrophysics, Foundation for Research and Technology-Hellas (FORTH), Heraklion, 70013, Greece}

\author[0000-0003-3265-4079]{Ryan Wollaeger}
\affiliation{Center for Theoretical Astrophysics, Los Alamos National Laboratory, Los Alamos, NM 87544, USA}
\affiliation{Department of Physics and Astronomy, Louisiana State University, Baton Rouge, LA 70803, USA}

\author[0000-0003-1087-2964]{Christopher J. Fontes}
\affiliation{Center for Theoretical Astrophysics, Los Alamos National Laboratory, Los Alamos, NM 87544, USA}



\begin{abstract}
In this paper, we introduce \texttt{SuperLite}, an open-source Monte Carlo radiation transport code designed to produce synthetic spectra for astrophysical transient phenomena affected by circumstellar interaction. \texttt{SuperLite} utilizes Monte Carlo methods for semi-implicit, semi-relativistic radiation transport in high-velocity shocked outflows, employing multi-group structured opacity calculations. The code enables rapid post-processing of hydrodynamic profiles to generate high-quality spectra that can be compared with observations of transient events, including superluminous supernovae, pulsational pair-instability supernovae, and other peculiar transients. We present the methods employed in \texttt{SuperLite} and compare the code's performance to that of other radiative transport codes, such as \texttt{SuperNu} and CMFGEN. We show that \texttt{SuperLite} has successfully passed standard Monte Carlo radiation transport tests and can reproduce spectra of typical supernovae of Type Ia, Type IIP and Type IIn.


\end{abstract}

\keywords{methods: numerical --- radiative transfer --- supernovae: general --- stars: evolution --- circumstellar matter}


\section{Introduction} \label{sec:intro}

Classical supernovae (SNe) are broadly divided into Type I, Type II and their sub-types based on their observed light curves and spectra following the explosion \citep[see][for a review of the SN classification system]{Filippenko:1997aa}. 
Unambiguous identification of the progenitor star is not possible for most of these events due to observational limitations. However, a significant progress has been made during the past several decades in understanding the mechanisms that drive the formation of these explosive events through theoretical modeling and the supporting observational evidence \citep[see, e.g.,][]{Branch:2017aa,Burrows:2021aa,Hillebrandt:2013aa}.
In the case of Type Ia supernovae (SNe Ia), the pre-explosion progenitor star has not been successfully observed in the archival images, with only one exception in the case of spectroscopically abnormal SN 2012Z \citep{Wright:2016aa}. The consensus about the progenitor of SN Ia being a white dwarf is based on the observations of early light curves that imply a compact star. The other widely accepted theory is that the progenitor white dwarf is part of a binary system, and it either accretes mass from its non-degenerate companion or merges with a white dwarf companion, which leads to the explosive core-carbon burning that results into a SN Ia. However, the evidence for the nature of such a companion star is also not well-established through observations \citep{Branch:2017aa}. There are several theories proposed to explain the explosion mechanism ranging from pure detonation to pure deflagration to deflagration-to-detonation models \citep{Hillebrandt:2013aa}. 

Even in the case of more commonly observed core-collapse supernovae (CCSNe or SNe Type II), the pre-explosion archival observations are limited to nearby events \citep[$d\leq 30 Mpc$,][]{Smartt:2009aa,2023MNRAS.519..471V}. The SNe II result from the gravitational collapse of the core of massive stars \citep[M $>$ 8\msun \ on the zero-age main-sequence, ZAMS, at solar metalicity,][]{Woosley:2002aa}. \cite{Bethe:1985aa} proposed a \textit{delayed neutrino heating} model in competition with then favored \textit{core bounce-shock} model of \citet{Colgate:1966aa} to explain these the CCSN mechanism. The neutrino-driven model of \citeauthor{Bethe:1985aa} was observationally confirmed when SN 1987A, identified as SN II, exploded in the Large Magellanic Cloudy (LMC) at a distance of less than 50 kpc \citep{Arnett:1989aa}. Kamiokande-II \citep{Hirata:1987aa}, Irvine-Michigan-Brookhaven \citep{Bionta:1987aa} detectors and the Baksan Scintillator Telescope \citep{Alexeyev:1988aa} observed the neutrino flux from the location of SN 1987A, which agreed with that predicted by the theory. Similar to SNe II, the theory also predicts that a flux of neutrinos can be observed from SNe Ia; however, it is about 4 magnitude smaller than SNe II. Theoretical calculations done for the detonation-to-deflagration model show that a SN Ia at a distance of about 10 kpc will be barely observable by the largest current and next-generation neutrino detectors \citep{Wright:2016aa}. Thus to make such an observation, a SN Ia has to explode at a close distance.

Inspite of the advancements in our knowledge over past several decades, there still exists a gap in our understanding of the characteristics of progenitor stars and the differences in the observed characteristics of SNe of the same type that result from these stars.
For example, the archival observations of the pre-explosion sites of CCSNe stipulate that the progenitor stars of SNe II are red supergiant stars (RSGs) that retained large hydrogen envelopes at the time of their explosion \citep{Smartt:2009ab,Smartt:2015aa}. On the other hand, the pre-explosion archival images indicate that SN 1987A resulted from an explosion of a blue supergiant star (BSG) that retained a substantial amount of hydrogen in its envelope \citep{Arnett:1989aa}. \cite{Arnett:1991aa} originally proposed a model with a ZAMS mass of $\sim$20 \msun \ and metallicity one-forth of the Sun with no mass-loss to explain a BSG progenitor star. \cite{Woosley:1988aa} and \cite{Woosley:1988ab} showed that the same model star with no convective overshoot or semi-convection remains a RSG through helium burning and later becomes a BSG to explode as a SN II. Alternately, the more recent binary merger models explain most of the observed properties of the progenitor of SN 1987A \citep{Menon:2017aa}. This demonstrates that the properties of the progenitor stars and the characteristics of the resulting SN depend greatly upon the choices of the initial conditions on the mass, metallicity and chemical composition, internal mixing, stellar rotation, mass loss in stellar wind, binarity, etc. \citep[some of these properties are explored in the four-article series,][]{Wagle:2019aa,Wagle:2020aa,Wagle:2020ab,Palani-Balaji:2022ws}.

There exist observed properties in some cases of SN explosions that are evidently the result of an interaction between the supernova outflow and the circumstellar material (CSM) that surrounds the progenitor star, formed by mass-loss prior to its explosion. In the absence of such interactions, the light curves are primarily powered by the initial shock energy deposited at the time of explosion and by radioactive decay of $^{56}$Ni or $^{56}$Co. However, in the presence of strong SN ejecta-circumstellar material interaction (hereafter, CSI), additional SN luminosity is displayed in the form of the ultra-violet (UV) or X-ray emission observed during the early phases to the radio emission observed during the late times. The nature of this emission depends upon the mass and location of the CSM and the properties and rate of expansion of the supernova ejecta, which in turn depends on the structure and evolution of the progenitor star. The CSI gives rise to a strong and fast shock wave in the CSM and a reverse shock in the SN ejecta. In such situations, the radiation leaks through the shocked region as the time-scale of photon diffusion is much shorter than the shock-crossing time, and the assumptions of local thermodynamic equilibrium (LTE) are no longer valid. The inverse-Compton scattering of photons due to the fast electrons in the shocked CSM produce the UV and X-ray emission. In addition the free-free radiation from both the forward- and reverse-shocked regions produce X-ray emission. The atoms in the high-density \textit{cool dense shell} (CDS) formed by the compressed shocked ejecta recombine to form the narrow-width emission lines observed in the SNe of Type IIn (``n" stands for narrow), especially the Balmer series line H$\alpha$. The wings of these emission lines can be broadened by multiple electron scattering. \citep[see][for further reading.]{Branch:2017aa,Chevalier:1994aa,Chevalier:2006aa,Chevalier:2017aa,2016MNRAS.458.2094D,2022A&A...660L...9D}

In addition to classical SNe IIn, over the last two decades, a new class of SNe that might indicate presence of strong CSI has been discovered \citep{Quimby:2011tc,Quimby:2012vh,Gal-Yam:2012uw}. These SNe exhibit luminosities of an order of magnitude higher than their classical counterparts. These eponymous superluminous supernovae (SLSNe) have 
observed properties, such as light curves (LCs) and spectra that cannot be solely attributed to any of the standard explosion mechanisms outlined above.
The SLSNe are broadly classified into two classes -- SLSN-I (hydrogen-poor) and SLSN-II (hydrogen-rich) -- to which most of the observed SLSNe belong. There is also a class of radioactively powered SLSNe (SLSN-R), which are less common, but better understood \citep{Gal-Yam:2012uw}. SN 2007bi is the first well-observed SLSN-R. The enormous luminosity of this SLSN implies large amount of radioactive nickel \citep[$>3$\msun,][]{Gal-Yam:2009aa} as expected from a full-fledged pair-instability supernova (PISN) model for very massive stars with initial mass well in excess of 100 \msun \ \citep{1967ApJ...148..803R,1983A&A...119...54E,2011ApJ...734..102K,2012ApJ...748...42C}.
One of the mechanisms among the plausible scenarios that could explain the observed luminosities of these SLSNe is the CSI \citep{Moriya:2018vh}.
The evidence of CSI has been observed in the case of luminous SN 2017hcc, which exhibited narrow emission lines in its spectra like a classical SN IIn. The late time multi-wavelength observations of this SNe show evidence of CSI \citep{Chandra:2022aa}. In rare cases, luminous SNe are observed to have SN Ia-like features before exhibiting SN IIn trends, which could be attributed to a WD exploding into dense CSM. \citep[e.g., SN 2018evt,][]{Yang:2023aa}

Understanding the nature of the episodic mass-loss associated with SNe and SLSNe is a challenge. Stars loose mass throughout their lives in the form of radiation-driven winds from their surface \citep{de-Jager:1988aa,Vink:2001aa}. The rate at which the mass is lost depends on the mass and luminosity of the star. The mass-loss is relatively unimportant in the pre-SN evolution of the low- and intermediate-mass stars ($\leq$10 \msun) until the final stages of evolution. However, in  higher-mass stars ($>$20 \msun ), a significant amount of the stellar mass can be taken away by the mass lost during stellar eruptions throughout their life. Therefore, the mass-loss is important for massive stars in determining the type of the resulting SN explosion. The CSM created by the moss-loss can come from both radiation-driven winds and episodic eruptions \citep{Smith:2014aa}. The mass-loss rates in the range of $10^{-4}$ to several \msun \ resulting in total CMS mass of $\sim$0.1 to tens of \msun \ have been inferred form observations \citep{Branch:2017aa}. The late-phase nuclear burning (especially Ne/O core-burning) in massive stars can lead to convectively- driven hydrodynamic waves (g-modes). Gravity waves and acoustic modes are excited at the interface of convective and radiative zones. These waves deposit super-Eddington heat near the surface, which can drive a appreciable mass-loss under certain conditions \citep{Quataert:2012aa,Fuller:2017aa,Fuller:2018aa}. Other mechanisms that can drive high mass loss involve binary interactions leading to common envelope ejection and binary mergers \citep{1984ApJ...280..771B,1989ApJ...337..849T,1991ApJ...373..246T}.

There are several existing and upcoming astrophysical transient search projects such as the James Webb Space Telescope (JWST) and the Wide Field Infrared Survey Telescope (WFIRST) that will most likely include primordial transient events in the early universe including the SN explosions of the first stars. In addition, facilities such as the Zwicky Transient Facility \citep[ZTF,][]{Bellm:2019aa} and the Vera Rubin Observatory \citep[formerly known as LSST; Large Synoptic Survey Telescope,][]{Ivezic:2008aa} will be discovering up to a million of new transient events every night. 

The prevalence of the role of CSI in many astrophysical transients necessitates the need for a radiation transport (RT) code that can simulate the emitted spectrum over the life of the transient under a variety of initial progenitor configurations. While a few codes that include the necessary assumptions to tackle the CSI regime exist, most of them are not fully accessible to the public and lack several components of physics that are important in strongly-shocked explosive outflows. In this paper, we introduce \texttt{SuperLite}, an open--source Monte Carlo Radiation Transport (MCRT) code that can be used to model interacting SN and transient spectra for easy comparison to observations.

In section \ref{sec:othercodes} we summarize the capabilities of other numerical frameworks that have been used to model interacting SN (Type IIn) spectra, in \ref{sec:methods}, we describe the numerical methods adapted in \texttt{SuperLite} and the modifications made in its parent code \texttt{SuperNu} to enable it to post-process the non-homologous SN ejecta. In section \ref{sec:verification}, we present results to verify that the code works as expected in the standard test case scenarios. In section \ref{sec:SN_applications}, we present the results of the spectra produced by \texttt{SuperLite} and compare it with other codes for several different SN types including the homologous and non-homologous velocity profiles. Finally, in section \ref{sec:results}, we discuss the results of the \texttt{SuperLite} code development along with the upcoming enhancements to the code. 

\section{Review of SN Radiation Transport Codes.} \label{sec:othercodes}

There are two distinct approaches to solving the equations of radiation transport -- the deterministic approach and the Monte Carlo approach. Both approaches have their advantages and disadvantages \citep[see][for further reading]{Castor:2004tn}. In the deterministic approach a full or partial set of radiative transport equations are solved by discretizing them into coupled system of algebraic equations. These methods can prove to be computationally very inefficient for large systems and have limited parallel scalability \citep{Abdikamalov:2012wg}.

In the Monte-Carlo (MC) methods, the radiation transport equations are not solved directly as is done with the deterministic methods. Instead, the trajectories of a number of packets of particles such as photons or neutrinos (referred to as MC particles or MC packets or MCP) are calculated stochastically using probability distribution functions (PDF) and pseudo-random numbers. Each MC particle represents a number of physical particles. Therefore, the number of MC particles required for calculations is much less than the number of physical particles they represent. The individual MC particles propagate seemingly randomly, but the ensemble of MC particles can provide an accurate representation of radiative transport process and the evolution of the radiation field. The biggest disadvantage of using MC methods is the stochastic fluctuations in the results of the MC calculations, due to its probabilistic nature. The MC methods suffer from MC noise, which roughly scales with the number of MC particles as $N^{-1/2}$ \citep{Abdikamalov:2012wg}. Thus one must use a large number of MC particles for better statistics, and hence, a better signal-to-noise ratio (SNR). 

Nevertheless, the MC particles can be decoupled and propagated independently, similar to the physical photons they represent. The number of MC particles is constrained by the computer memory and processing power, but with the modern high-performing computing systems this problem can be easily overcome with parallelization of the code. The computational cost increases further in the regions of high optical thickness. Special techniques need to be applied in the diffusion regime to overcome this issue \citep[see][for a review of advanced MC techniques]{Noebauer:2019aa}. A significant advantage of MC methods is that the scattering processes are easy to implement, compared to the deterministic methods discussed above. The propagation direction of the scattered particle can be chosen from a PDF constructed from the scattering kernels \citep{Abdikamalov:2012wg}. In addition, MC methods are less susceptible to numerical errors and are easy to extend to multidimensional time-dependent problems \citep{Noebauer:2019aa,Kasen:2006uw}. Therefore, the MC methods can be more advantageous for large and complicated systems, such as explosive and interacting outflows. 

The numerical frameworks that simulate the evolution or the radiation properties of the gas ejected during a SN explosion rely on several simplifying approximations; the most common being the assumption of homologous expansion of the ejected material ($v\propto r$), the LTE conditions, and the Sobolev approximation for line opacity \citep{Sobolev:1960aa}. (These approximations are discussed further in section \ref{sec:methods}).
For more common SN explosion types, such as SN Ia and SN IIP, these approximations yield LCs and spectra that are well in agreement with observations \citep{Branch:2017aa}. In the case of the CSI, most of these assumptions break down; more specifically that of homologous expansion of the ejecta and LTE. The Sobolev approximation also breaks down as a result of the non--homologous expansion of the shocked SN ejecta and CSM, since the line interaction surfaces are not parallel planes anymore \citep{Rybicki:1978aa}. To reliably produce the outcome in such cases, the hydrodynamic step has to be solved and the quantities need to be evaluated at each time step before the transport step is performed \citep{Roth:2015aa,Noebauer:2019aa}. 

There are codes that use numerical approaches to solve for the time-dependent non-equilibrium radiation-hydrodynamic (RHD) evolution of the SN ejecta, e.g., STELLA \citep{Blinnikov:1993aa,Blinnikov:1998aa,Blinnikov:2004aa,Blinnikov:2006aa}, which uses a radiation intensity moments scheme, or The SuperNova Explosion Code \citep[\texttt{SNEC},][]{Morozova:2015aa}, which uses flux-limited diffusion approximation. Both SNEC and STELLA are 1-D, multi-group RHD codes that evolve the radiation field to predict bolometric SN light curve. They calculate the ionization and excitation level populations of limited number of species in LTE conditions. SNEC imposes the same radiation and matter temperatures, while STELLA does not treat radiation in equilibrium with matter.
Codes such as SEDONA \citep{Kasen:2006uw} use MCRT methods to produces SN light curves, spectra, and polarization. SEDONA is a multidimensional, multi-frequency code that uses expansion opacity approximation. It also assumes that the ejecta are in homologous expansion.
\citet{Roth:2015aa} have coupled the MCRT code to 1-D, non-relativistic RHD solvers. These codes rely on simplifications, such as a single grey opacity or a limited number of groups, to speed up the calculations. They are useful in providing the general emission properties and model light curves, but they cannot predict a resolved spectrum that includes the effects of line emissivity and opacity. 
Some of the other popular publicly available RHD codes include \texttt{HERACLES} \citep{Gonzalez:2007aa}, \texttt{FLASH} \citep{Dubey:2012aa,Fryxell:2000aa}, and \texttt{CASTRO} \citep{Zhang:2011aa,Zhang:2013aa}.

There exist codes that predict a synthetic spectrum by solving the radiation transfer equations such as \texttt{CMFGEN} \citep{Hillier:1998vd,Hillier:2012vr}, \texttt{SYN++} \citep{Thomas:2011aa}, or by using MC techniques like in the case of \texttt{TARDIS} \citep{Kerzendorf:2014tc}, \texttt{SuperNu} \citep{Wollaeger:2013aa,Wollaeger:2014aa}, and \texttt{PHOENIX} \citep{Hauschildt:1992wo,Hauschildt:1999vw,Hauschildt:2004vm,van-Rossum:2012ur}.
\texttt{TARDIS} and \texttt{CMFGEN} are 1-D, MC codes that post-process the hydrodynamic profiles, and they are computationally inexpensive compared to the RHD codes that involve time-dependent calculations. \texttt{SYN++} is also a 1-D code that is a modern \texttt{C++} version of parametrized spectral synthesis code \texttt{SYNOW} \citep[\textit{Synthesis Now},][]{Branch:2009aa,Fisher:2000aa} that can be used for rapid analysis of SN spectra.  \texttt{SuperNu} is a time evolution code that advances the radiation field in each iteration. \texttt{TARDIS}, \texttt{SYN++}, and \texttt{SuperNu} all assume homologous expansion of the ejecta while implementing the radiation transport and for opacity and Doppler shift calculations. A comparison of RT codes performed by a collaboration of ten groups around the world that are developing existing RT codes is discussed in great detail by \citet[comprehensive supernova radiative-transfer code-comparison initiative; StaNdaRT,][]{Blondin:2022aa}.
The \texttt{TARDIS} code assumes that the effective photosphere is external to the volume 
in which a majority of luminosity is generated, which only holds true for early epochs of SNe Ia \citep{Kerzendorf:2014tc}. 

Table~\ref{tab:code_comp} summarizes the capabilities of some of the most frequently cited codes that have been used to model SN spectra and of our new, open--source code \texttt{SuperLite} that we introduce in this paper.

\begin{deluxetable*}{ccccccccc}
\tablecaption{Physics and approximations in some commonly used SN Radiation Transport codes. \label{tab:code_comp}}
\tabletypesize{\footnotesize}
\tablewidth{0pt}
\tablehead{
\colhead{Code} & \colhead{RT} & \colhead{post-} & \colhead{Homologous} & \colhead{Ionization \&} & \colhead{Line} & \colhead{Geometry} & \colhead{Parallel} & \colhead{Open} \\[-.2cm]
\colhead{} & \colhead{Method} & \colhead{processing} & \colhead{Expansion} & \colhead{Excitation} & \colhead{Opacity, $\kappa_{\nu}$} & \colhead{} & \colhead{Processing} & \colhead{Source}
}
\decimalcolnumbers
\startdata
\texttt{SuperLite} & MC & Yes & No & LTE(T$_e$) & CMF $\kappa_{\nu}$ & 1-D & MPI/OpenMP & Yes  \\
\texttt{SuperNu} & MC & No & Yes & LTE(T$_e$) & CMF $\kappa_{\nu}$ & multi-D & MPI/OpenMP & Yes \\
\texttt{CMFGEN} & RTE - CMF & Yes & No & \textit{dn/dt} & CMF $\kappa_{\nu}$ & 1-D & -- & Partly \\
\texttt{TARDIS} & MC & Yes & Yes & dilute-LTE(T$_r$) & Sobolev & 1-D & -- & Partly\\
\texttt{SEDONA} & MC & No & Yes & LTE(T$_e$) & Expansion & multi-D & MPI/OpenMP & Partly  \\
\texttt{STELLA} & RH-MG & No & No & LTE(T$_e$) & Expansion & 1-D & -- & Partly \\
\texttt{SYN++} & RTE - CMF  & Yes & Yes & LTE(T$_x$) & Sobolev & 1-D & MPI/OpenMP & Yes
\enddata
\tablecomments{Column headings: (1) Name of the code. (2) The numerical method used radiative transfer equation -- MC: Monte Carlo, RTE-CMF: Radiation Transfer equation, Co-moving Frame, RH-MG: multi-group radiation hydrodynamics. (3) If the code post-processes the profiles or takes time-steps and evolve the profiles. (4) Whether or not the ejecta are assumed to be in homologous expansion ($v=rt$) in the code. (5)  The method used to calculate ionization and excitation level populations. In LTE treatment, the Boltzmann excitation formula is used for excitation level populations and Saha-Boltzmann equation is used for ionization stage population, with setting the temperature to either the electron temperature (T$_e$) or the radiation temperature (T$_r$). \texttt{SYN++} uses user-specified temperature for local thermodynamic equilibrium (LTE). \texttt{TARDIS} uses dilute-LTE treatment as an approximation to non-LTE (NLTE) by using a dilution factor \textit{W} to scale the excitation level population. In NLTE, the rate equations are solved using the time-dependence on level populations ($dn/dt$). In a steady-state case, $dn/dt$=0. Currently, we are testing the NLTE for bound-bound opacity for hydrogen lines in \texttt{SuperLite}, which will be part of our future publications. (6) Line opacity calculations can be performed line by line explicitly in the co-moving frame (CMF) or by using the Sobolev approximation or by using an approximate frequency-dependent `expansion' opacity. (7) Available geometry. The 2-D and 3-D geometries are not being completely implemented and tested for \texttt{SuperLite}, hence, these are not part of the current version. However, the code is easily extendible to higher dimensions, and it will be part of our future publications. (8) The optional Message-Passing Interface (MPI) and OpenMP framework for parallel processing to speed-up the calculations. (9) All of the codes presented here have a public version available, but some codes do not include all of the methods and physics implemented in the code in their public versions. For example, the nonhomology and/or the NLTE applications for \texttt{SEDONA}, \texttt{TARDIS}, and \texttt{CMFGEN} are present in the literature, but the code is not publicly available in its entirety.
} 
\end{deluxetable*}

\section{Numerical Methods} \label{sec:methods}

\texttt{SuperLite} is a multigroup radiative transport code that uses Implicit Monte Carlo \citep[IMC,][]{Fleck:1971vj} and discrete diffusion Monte Carlo \citep[DDMC,][]{Densmore:2007aa,Densmore:2008aa,Densmore:2012tf} methods to model radiation transport processes in explosive and interacting outflows.
IMC solves the radiation transport equations semi-implicitly by treating absorption and emission as instanteneous effective scattering as explained in \cite{Fleck:1971vj}. The DDMC method is used to optimize IMC where local cell optical depth is higher than the user-specified value, to speed-up the calculations by treating several low mean-free-path scattering events with single diffusion events \citep[see, for instance,][]{Abdikamalov:2012wg}.
\texttt{SuperLite} is developed by significantly modifying the ``parent" \texttt{SuperNu} code and by relaxing some of its original assumptions that do not hold for outflows affected by shocked regions due to CSI.
\texttt{SuperLite} is a time-independent code that uses an Eulerian grid for MCRT simulation. The MC particle properties are mapped into the lab frame and advanced by the lab-frame form of Equation~\eqref{eqn:IMC} \citep[see, for instance,][equation 1]{Wollaeger:2013aa}. The co-moving frame energy of the particle is conserved during each interaction.
The opacity calculations are performed in the co-moving frame, by discretizing opacity into groups via direct integration over co-moving wavelengths. 
As an input to the \texttt{SuperLite} simulation, the spatial coordinates, velocity, temperature, and mass/density profiles are derived from a hydrodynamic or a RHD simulation at any given time since the SN explosion from an external code (such as STELLA). The outflow structures obtained from RHD simulations have the benefit of possessing implicit time-dependence in the temperature and material profiles. The \texttt{SuperLite} simulation itself is done in steady--state and is therefore time-independent.
\texttt{SuperLite} assumes that the information about the radiation field is fully taken into account by the radiation temperature structure predicted by the RHD profiles that are used as input. Hence, the radiation energy deposited into the spatial grid cells due to the radioactive decays of $^{56}$Ni or $^{56}$Co is not implicitly re-added.
In effect, \texttt{SuperLite} is a two-temperature code. The radiation temperature, $T_r$ is assumed to equal to electron temperature, $T_e$, if $T_r$ is not known a priori.

\subsection{Nonhomologous expansion}\label{subsec:nonhomology}

In the absence of CSI, the SN ejecta outflows are in a state of near free-expansion after the initial shock-breakout phase. Hence, at any given time after the explosion, the radial distance of a particular layer of the ejecta from the center of the expansion can be found if the velocity profile at any other time during the expansion is known. In other words, the radial distance of a layer of SN ejecta at any given time `$t$' since the explosion, $r(t)$, can be given by a simple relationship, $r(t) =vt$. This relationship also simplifies other quantities such as the divergence of velocity at time \textit{t}, that becomes $\vec{\nabla}\cdot\vec{v}(t) = 3/t$, and the density decreases with time as $\rho(t)\propto t^{-3}$. This is the standard assumption of homology which is hard-wired into many SN spectral synthesis codes, as discussed in section \ref{sec:othercodes}. In the case of shocked ejecta due to CSI, the assumption of homologous expansion fails as the velocity profile is nonhomologous in the entire region between forward and reverse shocks. In \texttt{SuperLite}, the equations of radiation transport are used in their original form without assuming homologous expansion, as discussed in sections \ref{subsec:steady} and \ref{subsec:prop}. 

\subsection{Steady-state approximation} \label{subsec:steady}

The Implicit Monte Carlo (IMC) equation in comoving frame, with assumptions for semi-relativistic simplifications from \citet{Castor:2004tn,Wollaeger:2013aa}, is written as:

\begin{eqnarray} \label{eqn:IMC}
\frac{1}{c} \frac{DI_{\nu}}{Dt} + \hat{\Omega}\cdot\nabla I_{\nu} - \frac{\hat{\Omega}\cdot\nabla\vec{v}\cdot\hat{\Omega}}{c}\nu\frac{\partial I_{\nu}}{\partial\nu} +\frac{3\hat{\Omega}\cdot\nabla\vec{v}\cdot\hat{\Omega}}{c} I_{\nu} - \nonumber \\
\frac{1}{c}\hat{\Omega}\cdot\nabla\vec{v}\cdot\left(\textbf{I}-\hat{\Omega}\hat{\Omega}\right)\cdot\nabla_{\hat{\Omega}} I_{\nu} + \sigma I_{\nu} \nonumber \\
= \frac{1}{4\pi}\frac{\sigma_e b}{\sigma_{P,e}} \left(1-f\right) \int\int \sigma' I_{\nu}' d\nu' d\Omega' + \frac{1}{4\pi}f\sigma bacT^4 \nonumber\\
+ \frac{1}{4\pi}\frac{\sigma_e b}{\sigma_{P,e}}(1-f)\dot{Q} \,,\quad
\end{eqnarray}
where \textit{c} is the speed of light, \textit{a} is the radiation constant, \textit{t} is time, $I_{\nu}$ is radiation intensity, $\hat{\Omega}$ is photon direction, $\nu$ is photon frequency, $\sigma$ is opacity (assumed to be only absorption here), $\sigma_e$ is the emission opacity, $\sigma_{P,e}$ is the Planck mean emission opacity, $\dot{Q}$ is a source term (for instance, a radioactive decay energy rate), \textit{b} is the normalized Planck function given by
\begin{equation}\label{eqn:norm_B}
b(\nu) = \frac{15}{\pi^4}\frac{h}{kT}\frac{(h\nu/kT)^3}{e^{h\nu/kT}-1},
\end{equation}
where, \textit{T} is temperature, 
and \textit{f} is the Fleck factor \citep{Fleck:1971vj}, given by 
\begin{equation}
f = \frac{1}{1+4aT^3c\sigma_{P,e}\Delta t/C_V},
\end{equation}
with $C_V$ being the material heat capacity at constant volume. Here $\Delta$t is the physical time-step size, which can be arbitrarily chosen as noted by \cite{Kerzendorf:2014tc}. 
The terms on the left-hand side of equation~\eqref{eqn:IMC} from left to right are the Lagrangian time derivative, spatial streaming operator, Doppler shift, adiabatic effect, directional aberration, and intensity attenuation from absorption, respectively \citep[see, for instance][]{Castor:2004tn}. The terms on the right-hand side are the effective scattering source (absorption instantly followed by emission), thermal emission (reduced by the amount treated with effective scattering), and any sources of energy in ejecta (again, radioactive energy), respectively \citep{Fleck:1971vj}.

In the steady state approximation, we can modify equation \eqref{eqn:IMC} by neglecting the local time derivative and setting $\Delta t \rightarrow \infty$, while preserving $O(v/c)$ effects. For infinite time-step size,
\begin{equation}
\lim\limits_{\Delta t \to \infty} f = 0.
\end{equation}

Under these assumptions, equation \eqref{eqn:IMC} becomes

\begin{eqnarray}\label{eqn:IMC_ss}
\frac{\vec{v}}{c}\cdot\nabla I_{\nu} + \hat{\Omega}\cdot\nabla I_{\nu} - \frac{\hat{\Omega}\cdot\nabla\vec{v}\cdot\hat{\Omega}}{c}\nu\frac{\partial I_{\nu}}{\partial\nu} +\frac{3\hat{\Omega}\cdot\nabla\vec{v}\cdot\hat{\Omega}}{c} I_{\nu} -  \nonumber \\
\frac{1}{c}\hat{\Omega}\cdot\nabla\vec{v}\cdot\left(\textbf{I}-\hat{\Omega}\hat{\Omega}\right)\cdot\nabla_{\hat{\Omega}} I_{\nu} + \sigma I_{\nu} \nonumber \\
= \frac{1}{4\pi}\frac{\sigma_e b}{\sigma_{P,e}} \int\int \sigma' I_{\nu}' d\nu' d\Omega'
+ \frac{1}{4\pi}\frac{\sigma_e b}{\sigma_{P,e}}\dot{Q}
\;\;\;\;. \quad
\end{eqnarray}
The right-hand side of equation \eqref{eqn:IMC_ss} implies that all photon collisions result in effective scattering, i.e., no MC packets are absorbed, hence conserving the co-moving frame energy of the packets as mentioned before. Each scattered particle is emitted with a new propagation direction. Effectively, all MC particles escape the domain as flux particles. For each particle that enters the inner boundary of the domain, a new particle with equal energy is generated with a random outward propagation direction at the inner boundary to conserve co-moving energy. For IMC, with these additional assumptions, the treatment of most of the terms remains the same; particle Lorentz transforms take care of $O(v/c)$ terms in the equation. The updated DDMC treatment is discussed in section \ref{subsec:prop}.

\subsection{Sourcing of MC particles}\label{subsec:sourcing}


The MC particles are sourced in each spatial cell according to the amount of radiation energy contained within the cell based on the input profile. As mentioned earlier, it is assumed that the energy deposition by radioactive decay has been already taken into account in the radiation field by the {\it hydrodynamics} code from which the profile is imported. Thus, the radioactive decay term is not separately added to the energy source for the particles.

In the TARDIS code \cite[][their Equation 10]{Kerzendorf:2014tc}, particles are sampled on an inner surface with radius $r_i$ and temperature $T_i$, and the energy per particle, for $N$ particles, is
\begin{equation}\label{eqn:source_ene}
E = \frac{4\pi r_i^2 acT_i^4}{4N} {\Delta}t.
\end{equation}
Assuming the radiation temperature is known, similar to equation \eqref{eqn:source_ene}, particles can be sourced in each spatial cell with energy $E_j$ at a surface with cell-centered radius $r_j$ given as,
\begin{equation}\label{eqn:source_ene1}
E_j = \frac{4\pi r_j^2 aT_{r,j}^4 V_j}{4N_j} \tilde{\Delta}t \,,
\end{equation}
where $T_{r,j}$ is the radiation temperature in cell $j$. Here, we have used $\tilde{\Delta}t$ to indicate individual particle's total propagation time to distinguish from the simulation time-step size $\Delta t$, that is taken to be infinite. The bolometric luminosity is computed as:
\begin{equation}
L = \frac{1}{\tilde{\Delta}t} \sum_p^{N_p} E_p,
\end{equation}
where \textit{p} is an index for escaped particles, $N_p$ is the total number of escaped particles, and $E_p$ is the energy weight at escape. Initializing particle weights with equation \eqref{eqn:source_ene1} and using the particle initialization sampling procedures for each cell ``surface” described, the result for \textit{L} is independent of $\tilde{\Delta}t$. However, this is only true under the assumption that particle propagation is independent of $\tilde{\Delta}t$ as well. To make \texttt{SuperLite} truly time-independent, we use the ``luminosity" weights, $L_j = E_j / \tilde{\Delta}t$ instead of equation \eqref{eqn:source_ene1} to instantiate the MC particles in each cell. This effectively eliminates the choice of arbitrary time-step size, which should not affect the results of the simulation. To ensure that the total output luminosity is equal to the user-input bolometric luminosity, the sum of the luminosity weights of all of the MC particles in all of the spatial cells is set to be equal to the user-input bolometric luminosity. As there is no loss of energy, the total output luminosity predicted by \texttt{SuperLite} remains unchanged.

The initial frequency of the particles is sampled with the normalized blackbody distribution given by equation \eqref{eqn:norm_B}. The particles are evenly distributed in the zone with the initial direction cosine $\mu = \sqrt{r}$, where $r\in(0,1]$ is a random number uniformly distributed in the interval from 0 to 1. 

\subsection{Particle propagation and Doppler correction}\label{subsec:prop}

The particle propagation is tracked through the transport (IMC) and diffusion (DDMC) steps. In the IMC step, the distance traveled by the packet, $d_{\rm p}$, is
\begin{equation} \label{eqn:dist}
d_{\rm p} = \min(d_{B},d_{thm},d_{col},d_{dop}) \,,
\end{equation}
where $d_B$ is the distance to the nearest cell boundary, $d_{thm}$ is the Thompon scattering distance, $d_{col}$ is the distance to effective scattering resulting from collision between a MC particle and an atom or an ion, and $d_{dop}$ is the distance to the Doppler shift to the adjacent upper or lower group, based on the divergence of velocity in that cell. 
As stated earlier, all collisions result in effective scattering. As the particles propagate, the position, frequency, and direction of propagation of the particle is updated.
In 1-D spherical geometry, the particle's radial coordinate, $r_p$ and the component of the direction of propagation along the radial direction, $\mu_p=\hat{r_p}\cdot\hat{\Omega}_p$ are updated as follows:
\begin{eqnarray} \label{eqn:r_mu}
\begin{array}{cc}
r_p' = \sqrt{r_p^2 + d_p^2 + 2\mu_p r_p d_p} \\
\mu_p' = (\mu_p r_p + d_p)/r_p' \,.
\end{array}
\end{eqnarray}
Here, $r_p'$ and $\mu_p'$ are particle's updated radial coordinate and the component of propagation direction along the radial coordiante, respectively.

The distances in equation \eqref{eqn:dist} are given by \citet[][spatial distance is used here instead of ``velocity" distance]{Wollaeger:2013aa}
\begin{eqnarray}
d_B = 
\left\{
\begin{array}{ll}
((r_{i-1/2}^2 - (1-\mu_p ^2) r_p^2)^{1/2} + \mu_p r_p) , \\
\quad \quad {\rm if} \ \mu < -\sqrt{1-(x_i/x)^2} \\
((r_{i+1/2}^2 - (1-\mu_p ^2) r_p^2)^{1/2} - \mu_p r_p) , \\
\quad \quad {\rm otherwise}
\end{array}
\right.
\end{eqnarray}
where, $r_{i-1/2}$ and $r_{i+1/2}$ are the left and right cell boundaries for a cell with index ``$i$", respectively (with $i\in[1...n]$, $n$ being the maximum number of spatial cells), and
\begin{equation}
d_{thm/col} = |\ln r| / \sigma_{thm/col} \,.
\end{equation}
Here, the form of the equation remains the same for the Thomson scattering distance and the distance to collision event. $\sigma_{thm}$ is the lab-frame Thomson scattering opacity and $\sigma_{col}$ contains the lab-frame opacity contributions detailed in section \ref{subsubsec:LTE}.
We generalize the formula for distance to Doppler shift, $d_{dop}$, between groups to non-homologous flows.  The equation to find this distance is derived from invariance of frequency in the $v=0$ frame (lab frame),
\begin{equation}
    \nu_p = \frac{\nu_p^{(0)}}{(1-\mu_pv(r_p)/c)} = \frac{\nu_{g\pm1/2}}{(1-\mu_p'v(r_p')/c)} \;\;,
\end{equation}
where $\mu_p'$ and $r_p'$ are given by equation \eqref{eqn:r_mu}.  We assume that within a spatial cell, the radial velocity is linear with respect to radius,
\begin{equation}
    v(r_p) = m_i (r_p - r_{i-1/2}) + v_{i-1/2} = m_ir_p + \tilde{v}_i \;\;,
\end{equation}
where $m_i=(v_{i+1/2}-v_{i-1/2})/(r_{i+1/2}-r_{i-1/2})$ is the slope of velocity across the radial domain of cell $i$ and $\tilde{v}_i = v_{i-1/2} - m_ir_{i-1/2}$.
We define a new inertial frame, the $\tilde{v}_i$ frame, to evaluated $d_{dop}$.
In this frame, the invariance is
\begin{equation}
     \tilde{\nu}_p = \frac{\nu_p^{(0)}}{(1-\tilde{\mu}_p(v(r_p)-\tilde{v}_i)/c)}
     = \frac{\nu_{g\pm1/2}}{(1-\tilde{\mu}_p'(v(r_p')-\tilde{v}_i)/c)} \;\;,
\end{equation}
where $\tilde{\nu}_p$ and $\tilde{\mu}_p$ are the frequency and radial direction component in the
new frame of reference, relating to the comoving frame, to order $v/c$, via
\begin{subequations} \label{eqn:tframe_trans}
    \begin{gather}
        \tilde{\nu}_p = \nu_p^{(0)}(1 + \mu_p^{(0)}(v(r_p)-\tilde{v}_i)/c) \;\;, \\
        \tilde{\mu}_p = \frac{\mu_p^{(0)} + (v(r_p)-\tilde{v}_i)/c}{1 + \mu_p^{(0)}(v(r_p)-\tilde{v}_i)/c}
        \;\;.
    \end{gather}
\end{subequations}
Similar to the lab frame, in the $\tilde{v}_i$ frame $\tilde{\mu}_p$ and $\tilde{\nu}_p$ are invariant.
Moreover, $\tilde{\mu}_p$ and $\tilde{\nu}_p$ can be found from the lab or comoving frame using Eqs. \eqref{eqn:tframe_trans}.
Furthermore, as in the lab frame, relating the initial and final $\tilde{v}_i$-frame direction components, 
$\tilde{\mu}_p'r_p' = \tilde{\mu}_pr_p + d_{dop}$, we find the distance to Doppler shift is
\begin{multline} \label{eqn:ddop}
    d_{dop} = \frac{c}{m_i}\left(1 - \frac{\nu_{g\pm1/2}}{\tilde{\nu}_p}\right) - \tilde{\mu}_pr_p \\
    = \frac{c}{m_i}\left(1 - \tilde{\mu}_p\frac{(v(r_p)-\tilde{v}_i)}{c} 
    - \frac{\nu_{g\pm1/2}}{\tilde{\nu}_p}\right) \\
    = \frac{c}{m_i\tilde{\nu}_p}\left(\nu_p^{(0)} - \nu_{g\pm1/2}\right)
    \;\;.
\end{multline}
We see that for homologous flow $m_i = 1/t$ and $\tilde{v}_i=0$, hence the $\tilde{v}_i$ frame becomes
the lab frame and Eq. \eqref{eqn:ddop} becomes the distance to Doppler shift presented by \cite{Wollaeger:2013aa}.
Importantly, this equation for Doppler shifting encodes redshifting based on the sign of the radial
velocity gradient, $m_i$, which can be seen by examining the right side of the third equality.
Given $\nu_{g+1/2} < \nu_p^{(0)} < \nu_{g-1/2}$ (decreasing group index implying increasing frequency),
if $m_i < 0$, then $\nu_{g\pm1/2} = \nu_{g-1/2}$ to give $d_{dop} > 0$, corresponding to blue-shifting.
Similarly, if $m_i > 0$, then $\nu_{g\pm1/2} = \nu_{g+1/2}$ to give $d_{dop} > 0$, corresponding to
red-shifting.
When $m_i=0$, there is no radial velocity differential across the cell, so the distance is infinite,
meaning the particle frequency does not shift toward either group boundary in the comoving frame. 

Since $\tilde{v}_i$ is artificial, it is possible that $|v(r_p)-\tilde{v}_i|$ violates the O(v/c)
approximation, and can even be greater than $c$.
In these instances, at least, a direct numerical solution for $d_{dop}$ can be obtained using
Newton-Raphson iteration, where the variable that is iteratively updated is $d_{dop}$.
We define the function $\varphi$ as follows,
\begin{equation}
    \label{eqn:nrfunc}
    \varphi(s) = m_i\mu_pr_p - c\left(1-\frac{\nu_{g\pm1/2}}{\nu_p}\right)
    + \tilde{v}_i\mu_p'(s) + m_is \;\;.
\end{equation}
By construction, $\varphi(d_{dop}) = 0$.
The derivative of Eq.~\eqref{eqn:nrfunc} is
\begin{equation}
    \label{eqn:dnrfuncds}
    \frac{d\varphi}{ds} = \tilde{v}_i\frac{d\mu_p'}{ds} + m_i
    = \tilde{v}_i\frac{(1-(\mu_p'(s))^2)}{r_p'(s)} + m_i \;\;,
\end{equation}
where use has been made of a standard identity for the path-derivative of $\mu_p(s)$.
Introducing a convergence tolerance $\varepsilon_{tol}$, the Newton-Raphson iteration
proceeds as follows.
\begin{enumerate}
    \item Estimate an initial value of $s=s_0$.
    \item While $|s_k-s_{k-1}| > s_k\varepsilon_{tol}$, for iteration $k$:
    \begin{enumerate}
        \item Evaluate $r_p'(s_k)$ and $\mu_p'(s_k)$ using Eqs~\eqref{eqn:r_mu} (replacing
        $d_p$ with $s_k$).
        \item Evaluate Eqs.~\eqref{eqn:nrfunc} and~\eqref{eqn:dnrfuncds} using $s=s_k$,
        $r_p'(s_k)$, and $\mu_p'(s_k)$.
        \item Calculate next iteration value:
        \begin{equation*}
            s_{k+1} = s_k - 
            \frac{\varphi(s_k)}{\displaystyle\left.\frac{d\varphi}{ds}\right|_{s=s_k}}
        \end{equation*}
        \item $k+1\rightarrow k$
    \end{enumerate}
    \item Set $d_{dop} = s_k$, where $s_k$ is the final iteration.
\end{enumerate}
We adopt this method to calculate the Doppler distance in \texttt{SuperLite}.

For DDMC, the energy loss due to Doppler shift is determined by the time-step size, which is infinite (or undefined) as discussed in section \ref{subsec:steady}. This choice would imply that the energy should be redshifted to 0, which is obviously incorrect. To best match the impact of detailed Lorentz transformations in the IMC portions, the DDMC particle’s time-of-flight in a spatial cell can be used, with energy loss begin determined by \cite[see, e.g,][]{Wollaeger:2013aa},
\begin{equation}
E' = E e^{-\nabla\cdot\vec{v_i}\cdot\delta t_k/3},
\end{equation}
where $\delta t_k$ is the time spent by the particle in cell \textit{j} on MC step \textit{k}. Note that the equation simplifies to the standard approach of \cite{Wollaeger:2014aa} when the flow is homologous, $\nabla\cdot\vec{v_j} = 3/t$ for all \textit{j}. To see this, if the time step $\Delta t$ is re-imposed,
\begin{equation}
\sum_k \delta t_k = \Delta t,
\end{equation}
so,
\begin{equation}
E_p^{(f)} = E_p^{(i)} \prod_k e^{-\delta t_k/t} =  E_p^{(i)} e^{-\Delta t/t} \,,
\end{equation}
where $E_p^{(i)}$ and $E_p^{(f)}$ are initial and end-of-time-step particle weight, respectively. For both IMC and DDMC, in steady state, non-Doppler adiabatic loss from expansion 
is assumed to have been taken into account by the pre-existing values of radiation temperature in each cell, $T_{r,j}$. 

The spatial leakage opacity formulation for DDMC follows \cite{Wollaeger:2014aa}, where optically thick
groups are regrouped (or collapsed) into single groups in order to optimize the simulations, by minimizing the direct treatment of line-to-line effective scattering.  However, we depart from previous DDMC formulations by developing a novel approach to Doppler shift from discontiguous groups in non-homologous flows.  Integrating the diffusion equation over the subset of groups to collapse into DDMC, assuming a Planck function Ansatz for frequency dependence within the resulting DDMC group, for a leakage opacity for leaking from the single DDMC group to one of several adjacent IMC groups via Doppler shift, we obtain
\begin{equation}
    \label{eq:dopleak}
    \sigma_{dop} = \begin{cases}
    \displaystyle\frac{\nabla\cdot\vec{v}_i}{3cb_{\mathcal{G}_{D}}}\sum_{g\in\delta\mathcal{G}_{D}^{+}}(\nu b)|_{g+1/2} \;\;\text{if }\nabla\cdot\vec{v}_i > 0 \;\;, \\\\
    \displaystyle\frac{-\nabla\cdot\vec{v}_i}{3cb_{\mathcal{G}_{D}}}\sum_{g\in\delta\mathcal{G}_{D}^{-}}(\nu b)|_{g-1/2} \;\;\text{if }\nabla\cdot\vec{v}_i < 0 \;\;,
    \end{cases}
\end{equation}
where subscript $\mathcal{G}_D$ is the subset of groups collapsed into one group for DDMC, denoting integration over the subset, $\delta\mathcal{G}_D^{\pm}$ is the subset of groups of the DDMC subset that are adjacent to IMC groups, either at the long (+) or short (-) wavelength side of the DDMC group.  If Doppler shift is sampled in DDMC from the discrete probability distribution formed by equation \eqref{eq:dopleak} and the other standard leakage opacities, then the particular group edge in $\delta\mathcal{G}_D^{\pm}$ that the particle escapes from is sampled from the discrete probability formed by the values $(\nu b)|_{g\pm1/2}$.

\subsection{Opacity and Emissivity Calculations} \label{subsec:opacity}
\subsubsection{Local Themodynamic Equilibrium} \label{subsubsec:LTE}

Thermodynamic Equilibrium (TE) describes an equilibrium state of interaction between gas and the radiation field -- a state where the radiation field is described by the Planck's law and the level populations by Saha-Boltzmann equations. Both of these are determined by the same state variable, the local temperature $T$, and the equilibrium state is described as the Local Thermodynamic Equilibrium (LTE). From microscopic point-of-view, in state of LTE, all atomic processes are in detailed balance, i.e., the number of atomic processes is balanced by the exact same number of inverse processes. 
We follow the same method as in \texttt{SuperNu} to determine the ionization balance and excitation level populations in each cell based on the LTE conditions. The Saha ionization equation \citep[see][]{Hubeny:2015wf} is:
\begin{eqnarray}\label{eqn:Saha}
\left(\frac{n_{0,j+1,k}}{n_{0,j,k}}\right)^* = \frac{2}{n_e} \left(\frac{2\pi m_e k_B T_e}{h^2}\right)^{3/2}\left(\frac{g_{0,j+1,k}}{g_{0,j,k}}\right) \nonumber \\
\exp\left(\frac{-\epsilon_{Ijl}}{k_B T_e}\right), \quad \quad
\end{eqnarray}
where, $n_{0,j,k}$ is the number density of the ground state population of an ion with net charge $j$ for a chemical species $k$. The asterisk `$*$' on the RHS of equation~\eqref{eqn:Saha} implies LTE. $n_e$ is the number density of the electrons, and $g_{0,j,k}$ is the statistical weight of the ground state of stage $j$ and $\epsilon_{Ijk}$ is the ionization energy of stage $j$. $T_e$ is the electron temperature and the other symbols have their usual meanings. 

The number density (originally ``occupation number") (or population) of a particular excited state $i$ in ionization stage $j$ can be determined using the Boltzmann excitation equation, once the ground-state populations have been calculated, i.e.
\begin{equation}\label{eqn:Boltzmann}
n^*_{ijk} = C_I \left(\frac{n_e n^*_{0,j,k}}{T^{3/2}}\right) \left(\frac{g_{ijk}}{g_{0,j+1,k}}\right) \exp\left(\frac{\epsilon_{Ijk} - \epsilon_{ijk}}{k_B T}\right),
\end{equation}
where, $\epsilon_{ijk}$ is the energy of excited level \textit{i}, and
\begin{equation}
C_I = \frac{1}{2}\left(\frac{h^2}{2\pi m_e k_B}\right)^{3/2} \,.
\end{equation}

By summing equation~\eqref{eqn:Boltzmann} over all levels within an ion stage, the total ion number density of successive ion stages can be obtained from the Saha-Boltzmann equation,
\begin{equation}\label{eqn:Saha_Boltzmann}
\left(\frac{N_{j,k}}{N_{j+1,k}}\right)^* = C_I \frac{n_e}{T^{3/2}} \left(\frac{U_{j,k}}{U_{j+1,k}}\right)^*\exp\left(\frac{\epsilon_{Ijk}}{k_B T}\right),
\end{equation}
where, $U^*_{j,k}$ is a LTE partition function. The following relationship between the ground state number density $n^*_{0,j,k}$ and the total number density $N_{jk}$ is also used:
\begin{equation}
N^*_{j,k} = \frac{n^*_{0,j,k}}{g_{0,j,k}} U^*_{j,k} \,.
\end{equation}
Equation \eqref{eqn:Saha_Boltzmann} is solved iteratively in \texttt{SuperLite} to determine the ionization populations and the electron density, until a desired accuracy is reached.

The line opacity is calculated, under the assumption of LTE, for the transition from a lower level \textit{l} to an upper level \textit{m} for within an ionization stage \textit{j} of chemical species \textit{k} (dropping the subscripts \textit{j} and \textit{k}) as
\begin{equation}\label{eqn:LTE_opacity}
\kappa^*_{l\rightarrow m} = n^*_l \left(\frac{\pi e^2}{m_e c}f_{l\rightarrow m}\right) \phi_{\nu} \left[1 - \exp \left(-\frac{\epsilon_{l\rightarrow m}}{k_B T}\right)\right],
\end{equation}
where, $f_{l\rightarrow m}$ is the oscillator strength and $\epsilon_{l\rightarrow m}$ is the corresponding transition energy. The symbol $\phi_{\nu}$ represents the line profile function, which is set to a delta function in \texttt{SuperLite} as in \texttt{SuperNu}. The factor in square brackets represents the effect of stimulated emission. The line emissivity is calculated under the LTE assumption from Kirchhoff's law:
\begin{equation}\label{eqn:LTE_emissivity}
\eta^*(\nu,T) = \kappa^*(\nu,T) \, B(\nu,T),
\end{equation}
where, $B(\nu,T)$ is the Planck function.
 
In \texttt{SuperLite}, the opacities are computed under the LTE assumption using a subgroup structure, as explained in \citet{Wollaeger:2014aa}. We include the atomic data for elements up to atomic number $Z = 30$. The opacity contributions include bound-bound (b-b), bound-free (b-f) and free-free (f-f) transitions, as well as standard elastic Thomson scattering opacity. The line opacities (b-b opacities) are calculated using the oscillator strength data for each atomic species using the Kurucz line list\footnote{http://kurucz.harvard.edu/atoms.html}. The data used in \texttt{SueprLite} includes hundreds of levels of each species with about 786,000 transitions in total. The b-f opacities are calculated using the analytic fit prescription described by \citet{Verner:1996wz} and the f-f opacities are calculated using the Gaunt factors based on the work of \citet{Sutherland:1998ww}. The total absorption opacity is the sum of the bound-bound, bound-free, and the free-free contributions for each subgroup. The wavelength groups have logarithmic spacing, while the subgroups have uniform linear spacing, unless otherwise specified. It is also possible to include a custom grid for the groups. The opacity mixing control parameter $\alpha_{\sigma} \in (0,1]$ introduced by \citet{Wollaeger:2014aa} is also included to linearly combine reciprocal (“Rosseland type”) and direct averages of opacity. The Planck-averaged opacity is calculated using the total absorption opacity. The Rosseland mean opacity includes the scattering contribution as well. However, the Rosseland mean opacity is calcualted only for output in the LTE version of the code.

\subsubsection{Non-Local Thermodynamic Equilibrium} \label{subsubsec:NLTE}
The assumption of statistical equilibrium holds true if the statistical timescale is much shorter than the timescale over which the radiation field changes. This is the case for the regions with high optical depths, but by definition not in the regions that play a role in shaping the spectrum. As the radiation leaves and reaches the observer from these regions, the assumption of LTE is violated. In non-LTE (NLTE) conditions, the detailed balance between excitation and de-excitation processes may be broken, leading to non-thermal populations of energy levels. To determine the level populations and ionization state of the gas, a complete set of rate equations has to solved \citep{Castor:2004tn,oelgoetz2007,Hubeny:2015wf}. In the matrix form, we can write these equations as
\begin{equation}
\textbf{A}\cdot\textbf{n} = \textbf{b}
\end{equation}
Here, the matrix \textbf{A}, which is often referred to as the \textit{rate matrix}, contains the \textit{radiative rates} ($R$, for interactions between particles and photons), and the \textit{collisional rates} ($C$, for interactions between two or more particles). The elements of matrix \textbf{A} for levels $i$ and $j$ are then:

\begin{eqnarray}
A_{ii} = \sum_{j\neq i}(R_{ij} + C_{ij}),\\
A_{ij} = -(R_{ij} + C_{ij}), \ for \ j\neq i
\end{eqnarray}

\textbf{n} is a vector of level populations, $\textbf{n} = (n1,n2,...,n_{NL})^T$, where $NL$ is the number of levels, and \textbf{b} is the vector of the rate at which a level's population changes, $dn/dt$. In steady-state, $dn/dt=0$.

In \texttt{SuperLite}, we have added the NLTE treatment for calculations of excited level populations of hydrogen up to the principle quantum number, n = 10. We use the \textit{quasistatic} approximation that the ionization balance can be decoupled from the excited-state population calculation. Furthermore, we assume that most of the population is in the ground state of the neutral and singly charged ion stages of hydrogen. The majority of the population flow between the two ion stages via ionization and recombination processes will occur through transitions that connect the ground states of the two ionization stages of hydrogen (rather than via processes that involve excited states), due to their massive populations relative to the excited-state populations. This dominant flow of population via ionization/recombination processes is, to a very good approximation, encapsulated within the prescribed ionization balance values defined by the LTE Saha ionization equation~\eqref{eqn:Saha}. Therefore, we do not consider the effect of ionization/recombination processes on the two ground states, but only on the processes that effect the excited-state populations. We solve the rate matrix to find the excited state populations of hydrogen, with an additional condition that the sum of ground- plus excited-state populations equals the ionization state population, $\sum{n_i} = N$. The first row of the \textbf{A} is replaced with 1's and the \textbf{b} is set to $(N,0,0,...,0)$, to reflect this boundary condition. To calculate the rate matrix we have included photo-ionisation and radiative recombination, and electron-impact collisional excitation and de-excitation rates calculated with the Los Alamos National Laboratory (LANL) suite of atomic physics codes \citep{LANL_suite}. The b-b line opacity and line emissivity are calculated using equations (14.12a) and (14.12b) of \citet{Hubeny:2015wf}. In our NLTE implementation, the b-b opacity for other elements are calculated under the LTE assumption, as described in section \ref{subsubsec:LTE} The b-f and the f-f opacities, and the Thomson scattering opacity are also calculated using the same methods as described in section~\ref{subsubsec:LTE}. The NLTE implementation is still in the testing phase, but some of the initial model results are displayed in section~\ref{sec:SN_applications}.


\section{Code Verification} \label{sec:verification}

\subsection{Line Transport in moving media - homologous and non-homologous cases} \label{subsec:line_trans}

\begin{figure*}[ht!]
\begin{center}
\gridline{\fig{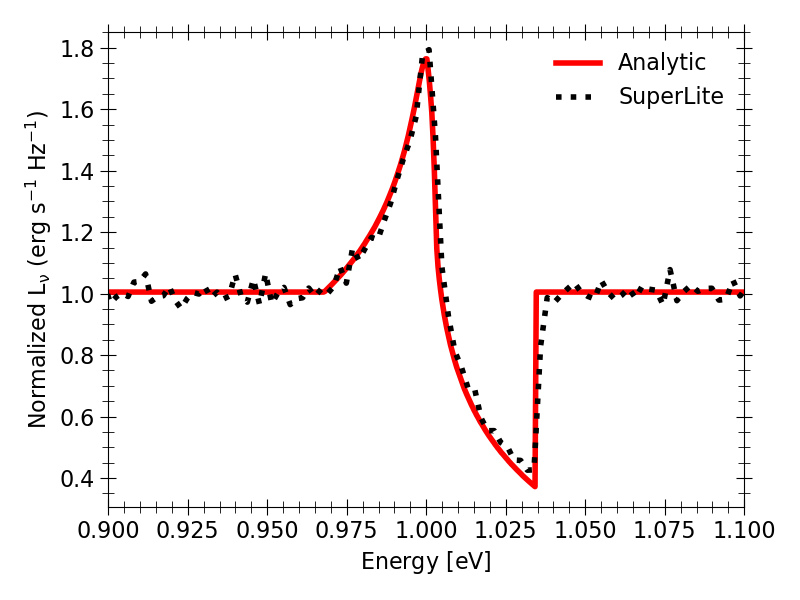}{0.45\textwidth}{(a)}
          \fig{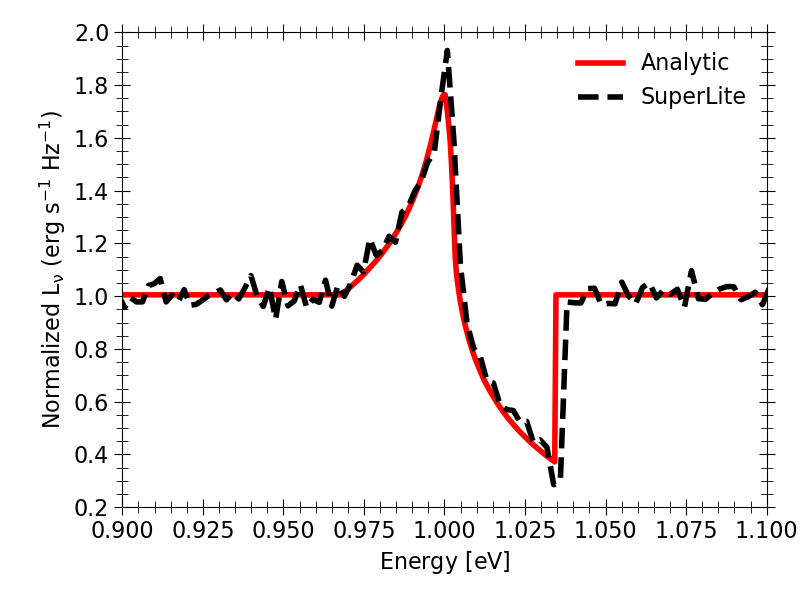}{0.45\textwidth}{(b)}}
\gridline{\fig{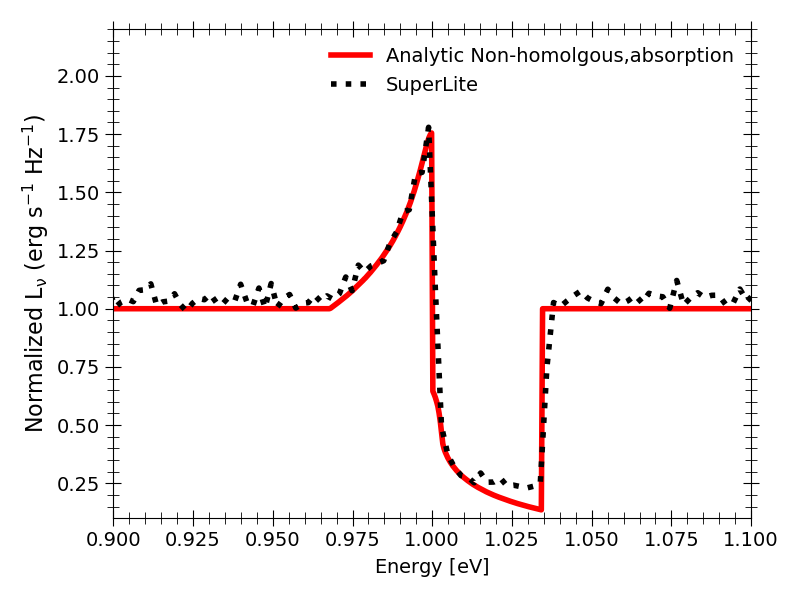}{0.45\textwidth}{(c)}}
\caption{The result of the test for line transport in a homologously expanding ejecta are shown in the left panel, while that for the non-homolgous case are shown in the right and bottom panels. Continuum photons are injected into a uniform sphere containing pure-scattering media with Sobolev line optical depth of $\tau_S = 1$. The line in the non-homologous region is scattering (right) or absorbing (bottom). Results from the MC transport are compared to the semi-analytic solution as discussed in section \ref{subsec:line_trans}. Within the uncertainty of MC noise, the modeled spectra compare well with the analytic solution. \label{fig:line_test}}
\end{center}
\end{figure*}

In this section, we present the test carried out to verify the \texttt{SuperLite} code using the transport of line radiation in a moving media as described in \citet[][section 7.2]{Roth:2015aa}. For this test, we consider an uniform density sphere of minimum radius, $r_{min} = 10^{14}$ cm and maximum radius, $r_{max} = 10^{15}$ cm, with uniform logarithmic grid of 100 spatial cells. The choice of density and temperature structures is not important for this problem. We choose an uniform density of $\rho = 10^{-11}$ g cm$^{-3}$ and uniform temperature $T_e = 10^3$ K. In this test case, we have considered the velocity profile to be homologous. For cell ``i", the velocity is given as $v_i = v_{max}(r_i/r_{max}) [cm/s]$, where $v_{max} = 10^8$ cm/s is adopted. We instantiate MC packets at the inner boundary of the domain with intensity 1 erg/cm$^{2}$/s/Hz/sr. (The choice of the intensity is arbitrary.) The MC packets are sampled with an uniform frequency distribution for this verification test. We use 2$^{19}$ MC packets for this test. A logarithmic frequency grid between the energies of 0.8 eV to 1.2 eV with 200 frequency groups is used. The line frequency of 1 eV is chosen so that the line is roughly in the middle of the frequency grid. The line opacity is calculated as 
\begin{eqnarray}
\kappa_{\nu} & = (dv/dr) * (\tau_S / c) * (\nu_c / d\nu_g) \nonumber \\ 
 & = (dv/dr)*(1/c)*(1/d\nu_g) [cm^{-1}]  
\end{eqnarray}
where, $\tau_S$ is the Sobolev optical depth, which was chosen to be 1 \citep[][equations 61]{Roth:2015aa}. $d\nu_g$ is the width of the frequency group containing the line centered at frequency $\nu_c =$ 1 eV. The semi-analytic solution to the emerging line profile is given by equation 60 of \citeauthor{Roth:2015aa}. As illustrated in Figure \ref{fig:line_test}a, the spectrum produced by \texttt{SuperLite} is in good agreement with the semi-analytic solution.

We also compare a version of the analytic solution to a non-homologous problem, where a wind profile with a constant velocity value of $v = 10^5$ cm/s is added to the homologous profile. The drop to $10^5$ cm/s introduces a single spatial cell with a negative velocity gradient.  We attempt to account for this in the semi-analytic solution by using the homologous intensity solution as an inner boundary source for the region with the negative velocity gradient. We test two variants of the non-homologous problem: one where the line in the negative gradient cell is purely scattering, as in the homologous region, and one where the line in the negative gradient cell is purely absorbing.  For the scattering variant, shown in Figure \ref{fig:line_test}b, a slight enhancement in the emission and absorption peaks of the P-Cygni profile is observed for the simulated spectrum, which can be attributed to part of the uncollided flux that is at slightly lower frequency than the absorption trough edge, which is getting blue-shifted back into resonance with the line. However, our analytic solution does not reflect this enhancement effect. For the absorption variant, the range of frequencies corresponding to the absorption trough happens to also be nearly equal to the range of lab frame frequencies that resonate with the line in the negative gradient cell.  The analytic solution for this is to simply attenuate the intensity by $e^{-\tau_S}$ over those frequency ranges.  As expected, for both the semi-analytic solution and the {\tt SuperLite} result, in Figure \ref{fig:line_test}c, only the absorption trough is enhanced.

\subsection{Type Ia SN spectra}\label{subsec:SN_Ia}

We verify the results of the snapshot post-processing with \texttt{SuperLite} by comparing with the results of radiation transfer calculations for time-dependent evolution of a SN-Ia by the \texttt{SuperNu} code. In the case of SNe Ia, the SN ejecta homologously expands following the shock break-out from the surface of a progenitor white dwarf that exceeds the Chandrasekhar mass. We use the 1D spherically-symmetric \texttt{w7} model of a carbon-oxygen white dwarf of about 1.37 \msun \ at the time of explosion \citep{Nomoto:1984vt}, that is included as a test case within the public version of the \texttt{SuperNu} code \citep{Wollaeger:2014aa} and has been used by several other authors \citep{Kasen:2006uw,Kromer:2009ug,van-Rossum:2012ur}. The input structure includes the density and abundance profile for several key isotopes up to $^{56}$Ni in a velocity grid.  The ground- and excited-level populations are calculated in \texttt{SuperNu} using the LTE assumptions discussed in section~\ref{subsubsec:LTE}. The model simulations start at 2~days from the time of explosion. The outer boundary of the ejecta outflow has a velocity of expansion about 0.07 times the speed of light. The output spectra is saved by \texttt{SuperNu} at the end of each time step. We adjusted the \texttt{SuperNu} code to enable the check-pointing of the SN ejecta profile at regular intervals, as the time-dependent MC calculations are advancing. The profile checkpoints include mass, electron temperature, and mass factions, which we used as inputs for the post-processing simulation with \texttt{SuperLite}.

For the purposes of this test, we chose to post-process a profile at 10 days since the time of explosion, near the time of peak luminosity. The radial coordinates are generated using the velocity grid with the relationship $r_i = v_i t$, as the expansion in this case is homologous. Here, ``$i$" is the cell index and ``$t$" is the time since explosion. 

\begin{figure*}[ht!]
\begin{center}
\includegraphics[width=0.9\textwidth]{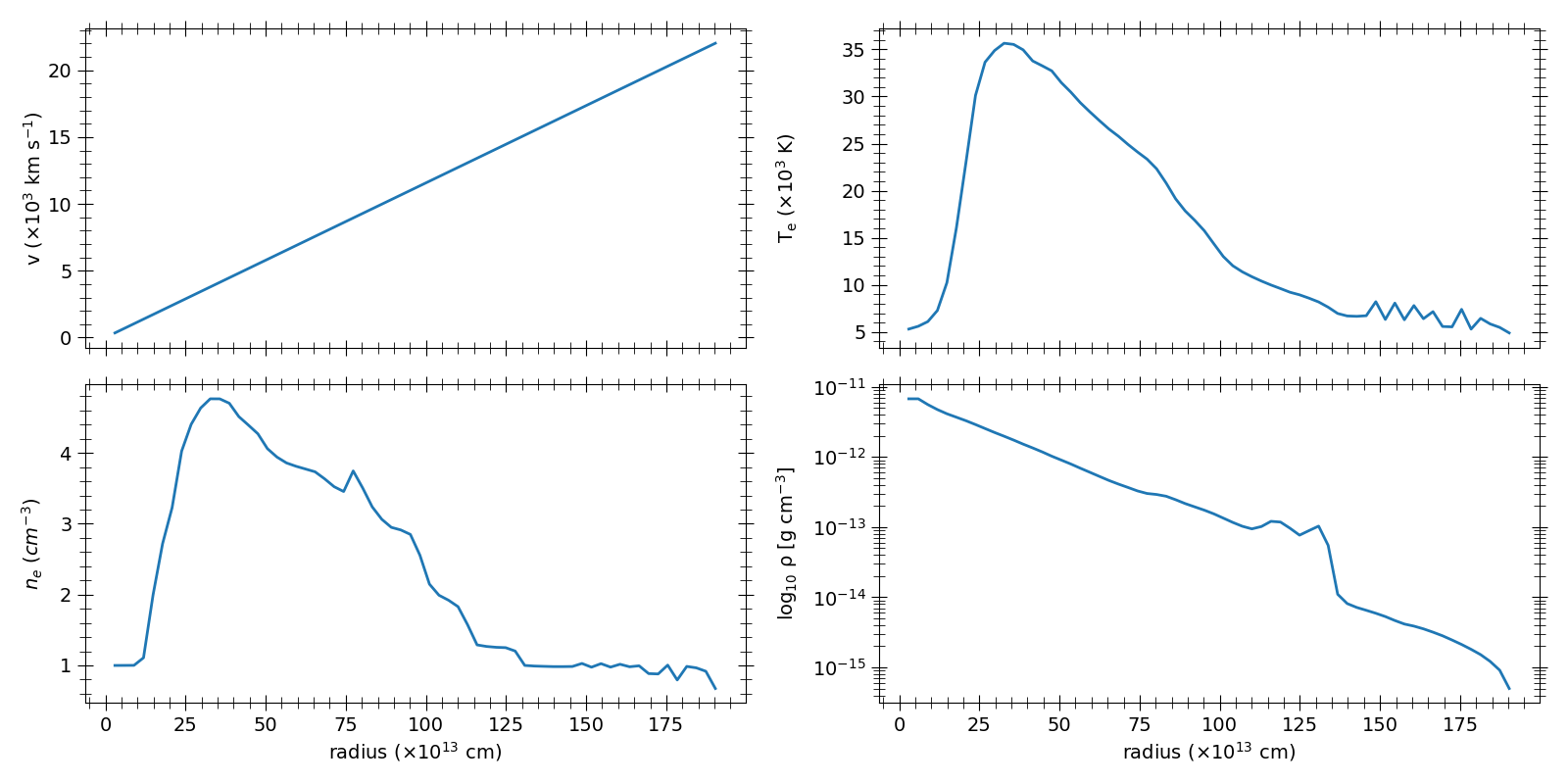}
\caption{The velocity, temperature, electron number density, and density profile plots at 10 days since the time of explosion for the {\tt W7} SN Ia model. 
\label{fig:w7_prof}}
\end{center}
\end{figure*}

The structure imported from \texttt{SuperNu} is shown in Figure~\ref{fig:w7_prof}. 
The bolometric luminosity predicted by \texttt{SuperNu} is used as an input to normalize the total input luminosity to the \texttt{SuperLite} simulation as described in section \ref{subsec:sourcing}.
The energy deposition due to radioactive decay processes are not included in \texttt{SuperLite}, as this energy is already accounted for in the \texttt{SuperNu} simulation. Similar to \texttt{SuperNu}, the opacity calculations are performed using the LTE assumptions. The electron temperature is used for both the sourcing of the MC packets and for the opacity calculations. The simulation input for \texttt{SuperLite} contains 64 spatial cells and 250 wavelength groups between the wavelength range of 1000 and 32000~\text{\AA}, both set to the same values as the \texttt{SuperNu} model simulation. About a million ($2^{20}$) MC packets were used for the \texttt{SuperLite} simulation, while for the \texttt{SuperNu} simulation about 16 million ($2^{24}$) initial and source MC packets each were used. \texttt{SuperLite} requires a smaller number of MC packets to obtain the same level of SNR. This is because, as a single-step, post-processing code with the steady-state assumption, all of the instantiated MC packets are processed through the ejecta. None of the MC packets are either absorbed (destroyed), censused, or buffered for the next time step in \texttt{SuperLite}.

\begin{figure*}[ht!]
\begin{center}
\gridline{\fig{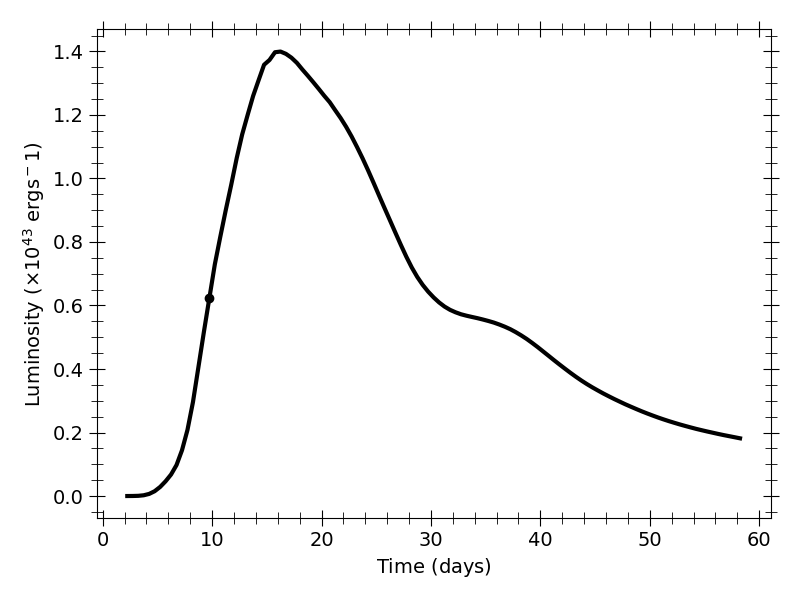}{0.45\textwidth}{(a)}
          \fig{spectra_w7_10d.png}{0.45\textwidth}{(b)}}
\caption{In left panel, the light curve predicted by \texttt{SuperNu} for the {\tt W7} SN Ia model is shown. The black dot marks the luminosity at 10 days since the time of explosion where profile was extracted from the SuperNu simulation for post-processing with SuperLite. In the right panel, a comparison of spectra at 10 days since the time of explosion produced by \texttt{SuperNu} and LTE version of \texttt{SuperLite} are shown. \label{fig:w7_10d}}
\end{center}
\end{figure*}

Figure~\ref{fig:w7_10d} shows a comparison between the spectra predicted by \texttt{SuperNu} at the time step when the output profile was saved at 10~days after explosion with the spectra predicted by the LTE version of \texttt{SuperLite} in right panel, along with the light curve for the SN generated by \texttt{SuperNu} in the left panel. 
Within the uncertainty of the MC noise, the spectra produced by both codes are in good agreement. The \texttt{SuperLite} simulation was completed within five minutes on a typical desktop with parallel processing using 6 MPI ranks. 

\subsection{Profile Truncation Depth}

\begin{figure*}[ht!]
\begin{center}
\gridline{\fig{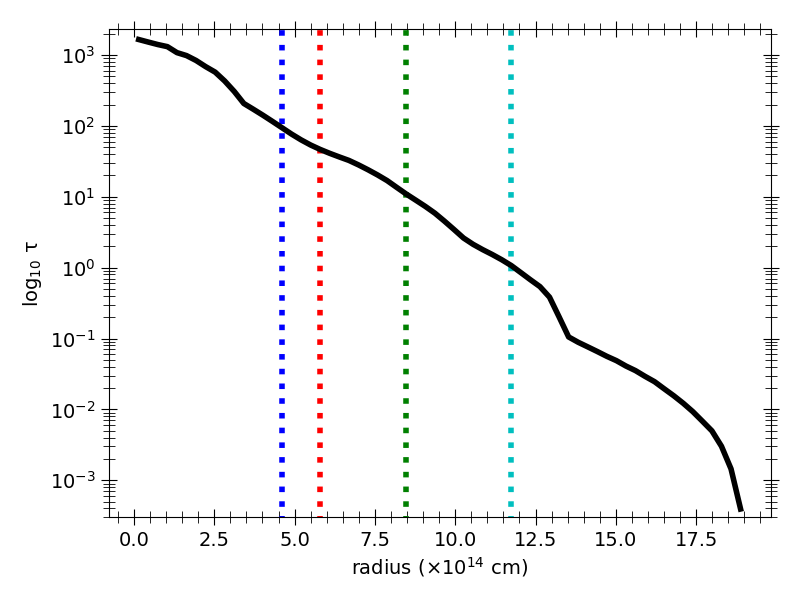}{0.45\textwidth}{(a)}
          \fig{spectra_w7_10d_trunc.png}{0.45\textwidth}{(b)}}
\caption{The effects of truncation depth of the ejecta on the predicted spectra are shown. In the left panel, the optical depth ($\tau$) calculated from the Rosseland mean opacity is shown for the ejecta profile of the 
\texttt{W7} SN Ia model. The optical depth is set to 0 at the location of the observer at infinity, so the optical depth increases into the ejecta. The vertical lines in the left panel mark the location of $\tau =$ 100, 50, 10, and 1, respectively, from left to right. In the right panel, a comparison of spectra produced by \texttt{SuperLite} by truncating the ejecta at different optical depths from the surface inward is shown. The black, solid line shows the spectrum for the whole ejecta profile. The spectrum remains mostly unaffected up to $\tau=$50 (red, dashed line). It changes slightly at longer wavelengths for $\tau=$10 (green, dashed-dotted line). However, it changes significantly when the ejecta profile is truncated at $\tau=$1 (cyan, thin solid line). \label{fig:w7_10d_trunc}}
\end{center}
\end{figure*}

In Figure \ref{fig:w7_10d_trunc}, we show the effect of truncation depth into the ejecta profile on the calculations of spectra by \texttt{SuperLite} for the \texttt{w7} model. As \texttt{SuperLite} is a post-processing code, it doesn't need entire ejecta to process the spectra. Including the line forming region is sufficient to predict the spectra within the uncertainties of MC calculations. To demonstrate this, we truncated the ejecta profile for the \texttt{w7} at different location based on the optical depth ($\tau$) calculated using the Rosseland mean optical depth, and simulated the spectrum using the \texttt{SuperLite} code for this model. The optical depth is set to 0 at the location of the observer at infinity. Thus, it increases into the ejecta. The left panel of the figure shows $\tau$ with the locations of $\tau =$ 100, 50, 10, and 1 marked with vertical lines. The ejecta profile was truncated from the surface inward at these locations into the ejecta. The right panel of the figure shows a comparison for the simulations with different truncation depths. As seen in this figure, the spectra predicted by the code remains mostly unaffected until about the truncation depth at $\tau =$ 10 (green, dashed-dotted line). The spectra for $\tau =$ 1  (thin cyan line) is significantly different than the original simulation with the entire ejecta profile (black, solid line). As the optical depth varies with the frequency of the photon packet, the depth in to the outflow profile up to the Rosseland mean optical depth of $\tau =$ 1 is not sufficient enough for correctly predicting all of the transitions. We recommend including the outflow profile up to a Rosseland mean optical depth of $\tau \approx$ 30--50 for better prediction of the synthetic spectra.

\section{Applications to supernova spectral modeling}\label{sec:SN_applications}

Here, we present the synthetic spectra produced with \texttt{SuperLite} post-processing simulations for various types of SNe including both the homologous and non-homologous expansion regimes, and compare them with observations or results from the CMFGEN code. 

\subsection{Type IIP SN spectra}\label{subsec:SN_IIP}

In this section, we present the results of \texttt{SuperLite} post-processing for two typical SN IIP models and compare the results with the time-evolution simulation performed with \texttt{SuperNu} in one case and to the observations of a typical SN IIP in the other case.

\subsubsection{Model s18.0} \label{subsubsec:s18.0}

For the first case, we use model s18.0 from the suite of SN IIP model simulations presented by \citet{Curtis:2021to}. The model consists of a star with ZAMS mass equal to 18~\msun and solar metallicity, and a pre-SN mass of 14.5~\msun \ at the time of explosion, with a 9.25~\msun \ hydrogen envelope. The pre-SN radius of the progenitor star is 1010~\rsun . An explosion energy of 1.45$\times 10^{51}$~erg is assumed. The \texttt{SuperNu} comparison model corresponds to 20 days from time of explosion. The velocity profile is shown in Figure~12 of \citep{Curtis:2021to}. We use the \texttt{SuperNu} input files kindly provided by Dr. Sanjana Curtis (private communication) to re-run the \texttt{SuperNu} simulation with the same setup as in their paper. The SuperNu model was run with $2^{18}$ initial MC packets and additional $2^{18}$ source packets were added every timestep, with a maximum limit of $2^{28}$ MC packets. The model contains 689 spatial cells, and 1000 wavelength groups with the wavelength range of 1000--32,000 \text{\AA}. Similar to \texttt{w7} model discussed earlier, we have used a modified version of \texttt{SuperNu} code that saves SN ejecta profiles at checkpoints every few days. The profiles from this \texttt{SuperNu} simulation are used for post-processing with \texttt{SuperLite}.

\begin{figure}[ht!]
\begin{center}
\includegraphics[width=0.45\textwidth]{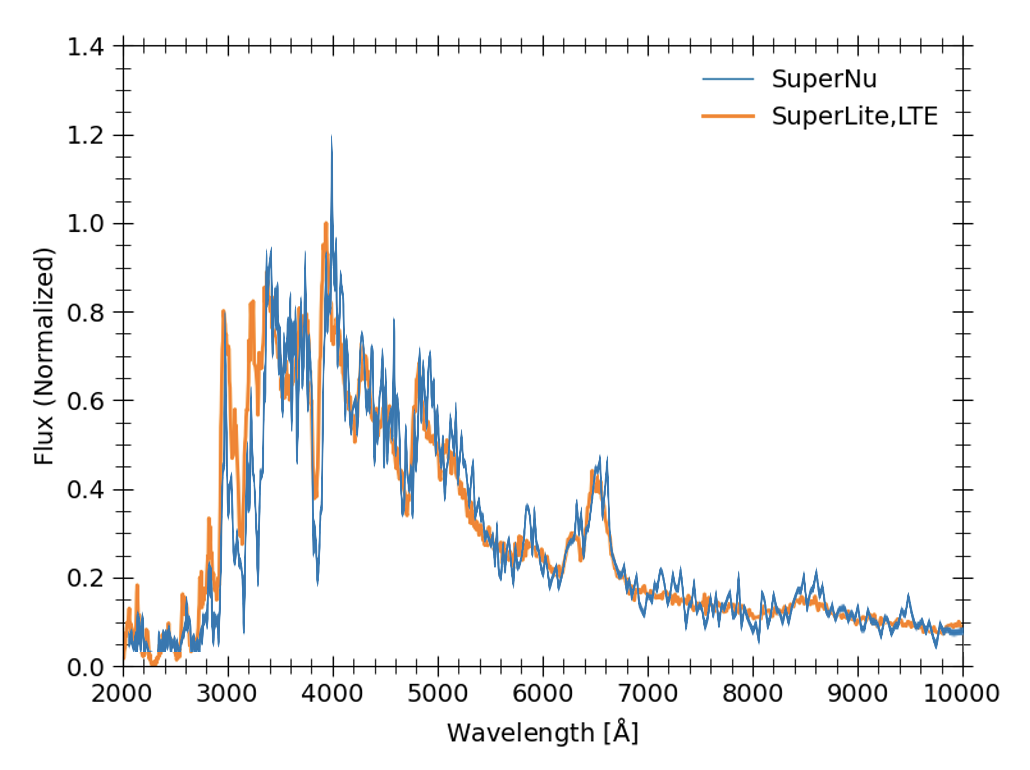}
\caption{A comparison of spectra produced with the \texttt{SuperNu} and the LTE version of the \texttt{SuperLite} codes for a Type IIP model at 20 days since the time of explosion is shown. The \texttt{SuperNu} spectrum is taken from figure 5 of \citet{Curtis:2021to}. The two spectra are in good agreement with each other. \label{fig:San_IIP_20d}}
\end{center}
\end{figure}

For a comparison of the \texttt{SuperNu} output spectra to that produced with \texttt{SuperLite} in 1D, we choose a phase corresponding to 20 days after the SN explosion. The hydrogen recombination front begins to recede into the H--rich SN ejecta at around ~20~days, marking the beginning of the eponymous plateau phase that lasts for about 104~days \citep[see figure 13 of][]{Curtis:2021to}. For the 1-D simulation calculated with \texttt{SuperLite}, we generate the radial coordinates under the assumption of homologous expansion, as in the previous case, because the \texttt{SuperNu} model simulation is run on a homologous velocity grid. The ejecta was truncated at the Rosseland mean optical depth of 100. The model contains 172 spatial cells, and 6000 wavelength groups with the wavelength range of 1000--30,000 \text{\AA}. This simulation was run with about a million ($2^{20}$) MC particles is the same as that used in the {\tt w7} model. The spectra generated with \texttt{SuperNu} and \texttt{SuperLite} are shown in Figure~\ref{fig:San_IIP_20d}. The spectrum generated with \texttt{SuperNu} from figure 5 of \citet{Curtis:2021to} is superimposed on the \texttt{SuperLite} spectrum. The \texttt{SuperNu} spectrum is smooted, while the \texttt{SuperLite} spectrum is not smoothed. The two spectra are in a good agreement with each other.

\subsubsection{A classical SN IIP: SN 1999em} \label{subsubsec:99em}

Here, we present a comparison between the synthetic spectra predicted by \texttt{SuperLite} to the observed spectra for a typical SN IIP, SN 1999em that exploded in the host galaxy NGC 1637 \citep[Lick Observatory Supernova Search][]{Li:1999aa}. Based on the pre-explosion observations of the progenitor star, \citet{Smartt:2002aa} derived an upper limit on of 12$\pm$1 \msun \ on the ZAMS mass of the progenitor star of SN 1999em, assuming a distance of 7.5$\pm$0.5 Mpc determined by the expanding photosphere method \citep{Hamuy:2001aa}. \citet{Smartt:2009ab} derived an upper limit of 15 \msun \ for the same, assuming a distance of 11.7$\pm$1.0 Mpc determined using the Cepheid variables in the host galaxy \citep{Leonard:2003aa}. \citet{Utrobin:2007aa} performed optical hydrodynamic model fitting to the bolometric light curve and spectroscopic evolution of H$\alpha$ line for SN 1999em to determine the progenitor properties. With an adopted distance of 11.7 Mpc, their model with solar metallicity and a mass of 19$\pm$1.2 \msun \ for the ejected envelope mass, a pre-SN radius of 500$\pm$200 \rsun \ exploded with 1.3$\pm$0.1 Bethe energy and 0.036$\pm$0.009 \msun \ of nickel mass fitted well with observations. For the shorter distance of 7.5 Mpc, their model with 16 \msun \ of envelope mass, pre-SN radius of 375 \rsun \ exploded with 0.686 Bethe energy and 0.0162 \msun \ of nickel mass was the best fit model. We do not intend to fit the observed SN 1999em spectra, but we use the example of SN 1999em to illustrate the capacity of our code to model a typical SN IIP spectrum. To this effect, we use one of the two models of SN 1999em-like progenitor star explored by \cite{Paxton:2018aa}, which we describe below. The properties of the model are listed in their table 3.

\begin{figure}[ht!]
\begin{center}
\includegraphics[width=0.45\textwidth]{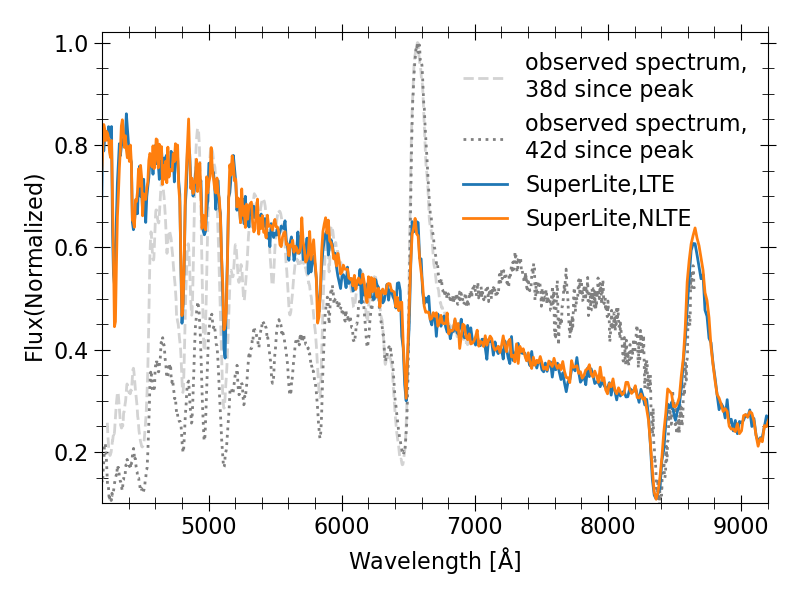}
\caption{A comparison of spectra produced with the the LTE and the NLTE versions of the \texttt{SuperLite} code for a SN IIP model for SN 1999em at 40 days since explosion. The NLTE implementation is still being tested. The observed spectra from \citet{Faran:2014aa,Leonard:2002aa} for SN 1999em is shown in the background for comparison. The lightgrey curve shows the observed spectra obtained at 38d and the grey curve shows the observed spectra obtained at 40d since the peak luminosity. \label{fig:99em_40d}}
\end{center}
\end{figure}

We evolve a progenitor star with solar metallicity and ZAMS mass of 19 \msun \ with the stellar evolution code MESA \citep[Modules for Experiments in Stellar Astrophysics;][]{Paxton:2011aa,Paxton:2013aa,Paxton:2015aa,Paxton:2018aa,Paxton:2019aa}. This progenitor star has a pre-SN mass of 17.8 \msun , with a massive H-rich envelope and pre-SN radius of 603 \rsun . The model star is exploded with STELLA (a version of the code provided with the MESA distribution), by injecting energy of 1 Bethe at the base of the envelope and assuming nickel mass of 0.042 \msun . The profiles are generated at several checkpoints by STELLA. These profiles include the electron and radiation temperatures, as well as the total bolometric luminosity estimated by STELLA. These profiles are used for post-processing with \texttt{SuperLite} to generate synthetic spectra. For the \texttt{SuperLite} simulation, we choose a phase corresponding to 40 days since the maximum bolometric luminosity of the SN. This phase is within the plateau of the SN light curve. We truncate the ejecta profile at the STELLA-predicted optical depth of 100 for the \texttt{SuperLite} simulation. The input structure contains 154 spatial cells. We choose 6000 wavelength groups with a range of 1 to 30,000 \text{\AA}. About a million ($2^{20}$) MC particles are instantiated for the simulation. Figure \ref{fig:99em_40d} shows the spectra predicted by the LTE and NLTE versions of the \texttt{SuperLite} code. The observed spectra taken at 38 and 42 days since the peak in the light curve of the SN is plotted in the figure for comparison with \texttt{SuperLite} prediction. The observed spectroscopic data is obtained using 1m Nickel reflector at Lick Observatory \citep{Faran:2014aa,Leonard:2002aa}. The \texttt{SuperLite} simulation qualitatively predicts the observed line features with the P-Cygni profiles, with blue-shifted emission components, and the continuum in that taken at day 38. However, this observed spectrum extends only to wavelength of 6800 \text{\AA}. The Ca II feature around 8500 \text{\AA} in the spectrum observed at day 42 are also well-predicted by \texttt{SuperLite} code. The observed spectrum at day 42 is not corrected for the effects of dust absorption, while the model spectra does not take these effects into account. This explains the differences in the continuum in the two spectra at the shorter wavelengths.

\subsection{Type IIn SN spectra}\label{subsec:SN_IIn}

One of the main motivations behind the development of \texttt{SuperLite} was to model the spectra of astrophysical transients that are affected by strong interaction with a CSM. Toward that end, in this section we present the first interacting SN spectra simulated with our code.

\subsubsection{A classical SN IIn} \label{subsubsec:A4}

\begin{figure*}[ht!]
\gridline{\fig{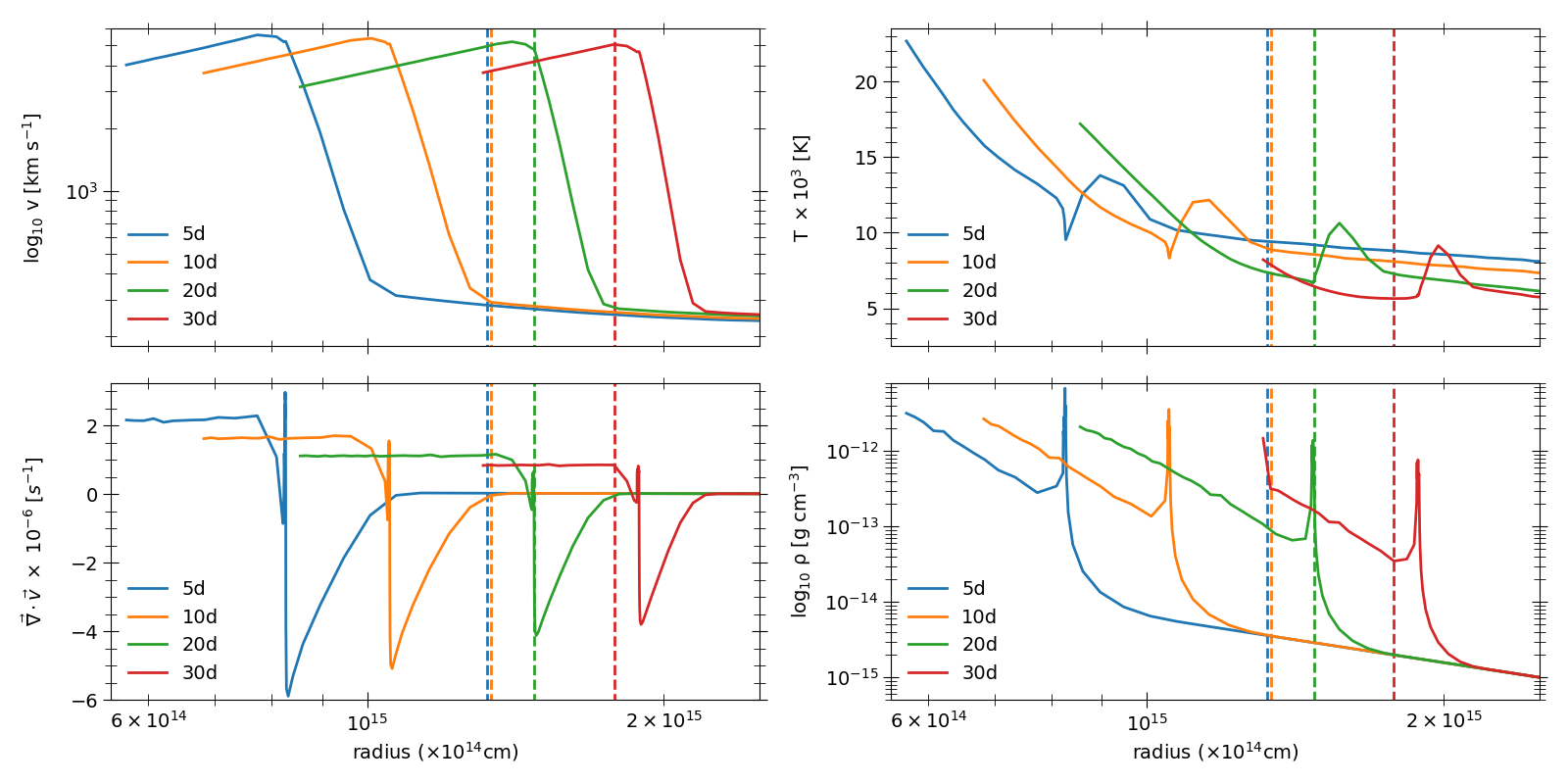}{0.95\textwidth}{(a)}}
\gridline{\fig{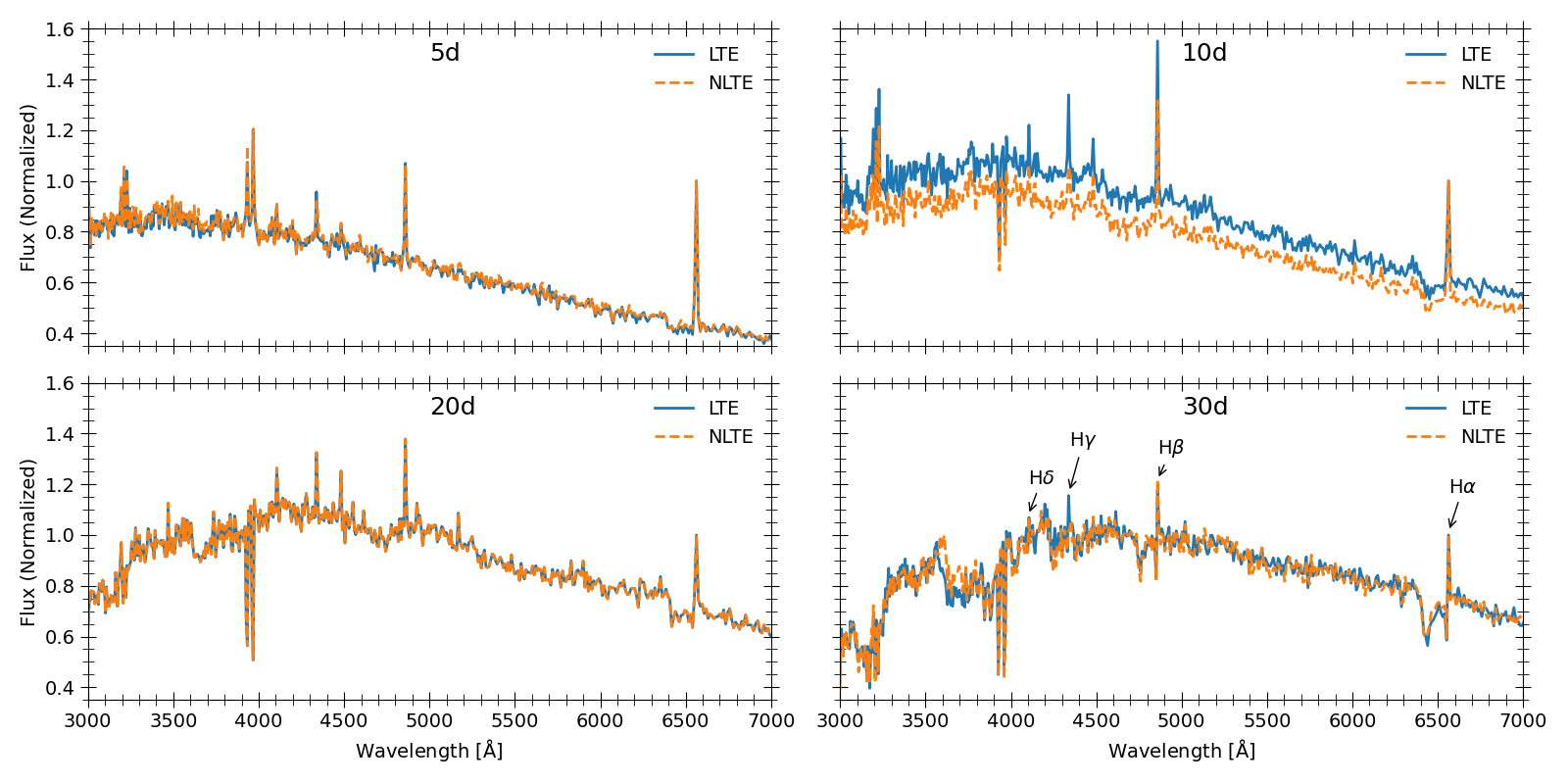}{0.95\textwidth}{(b)}}
\caption{In panel (a), the velocity, electron temperature, electron density, and density profiles at 5, 10, 15, and 20~days from the time of explosion for a SN IIn model A4 are shown. The vertical dashed lines in the velocity plot mark the radial coordinate of the cell where the Rosseland mean optical depth is 1. As the ejecta expands and cools down, the photosphere recedes into the shocked region. In panel(b), the spectra produced with the LTE and NLTE versions of \texttt{SuperLite} corresponding to the same phases are shown. The spectra are normalized to H$\alpha$ line strength for ease of comparison. The NLTE version of the code is still being tested. \label{fig:A4_evol}}
\end{figure*}

We construct a model progenitor star for a typical SN IIn with ZAMS mass of 19 \msun \ with the same specifications as described in section \ref{subsubsec:99em} for the SN 1999em. In addition, we append a dense H-rich CSM of 0.2 \msun \ around the progenitor star that corresponds to a mass-loss rate of 0.025 \msun /yr for 8 years before the core-collapse stage, with the velocity of the stellar wind set to 200~km\,s$^{-1}$. A power law index of 2 is used to construct the CSM density profile around the progenitor star. The model progenitor star is then exploded into the surrounding CSM using STELLA. The explosion energy of 0.78 Bethe is injected at the base of the envelope. A total nickel mass of 0.042~\msun \ is used for this simulation. The non-homologous expansion of the SN ejecta is tracked by the time-dependent RHD evolution. As earlier, the profiles generated at several checkpoints by STELLA that include the electron and radiation temperatures, as well as the total bolometric luminosity are used for the \texttt{SuperLite} simulation.
We refer to this model as A4, and it belongs to a suite of models that we have explored with STELLA for the interacting SNe. The complete suite of models along with the post-processed \texttt{SuperLite} simulations will be presented in a future publication. The profiles generated by STELLA at 5, 10, 20, and 30 days since the maximum in the bolometric luminosity are used to run the post-processing model simulations with the \texttt{SuperLite} code to generate spectra for the A4 model. In the top 4 panels of Figure~\ref{fig:A4_evol}, velocity, temperature, density, and velocity gradient profiles that we use as an input for our post-processing simulations with \texttt{SuperLite} are shown for each of these phases. The vertical lines show the location of the Rosseland mean optical depth of 1. The optical depth is set to 0 at the location of the observer at infinity, so the optical depth increases in to the profile. As the time progresses, the photosphere recedes deeper traversing the shocked region around 30 days. The bottom 4 panels in Figure~\ref{fig:A4_evol} show the spectral evolution of the SN ejecta for the A4 model at these selected phases. The strong and narrow emission lines, especially the Balmer series lines H$\alpha$ and H$\beta$ can be seen in the spectra. As the ejecta expands and cools down, the continuum shifts toward longer wavelengths. The spectra resulting from our experimental NLTE implementation has also been shown in this figure. The LTE and NLTE spectra look very similar until 30 days phase, where the photosphere is traversing the shocked region, where the LTE assumption breaks.

\subsubsection{A Luminous SN IIn - SN 2017hcc}

We simulated another example of a SN IIn that undergoes stronger CSI. This model is similar to the progenitor of SN 2017hcc that exploded in a spiral dwarf galaxy of near-solar metallicity at a distance of 73 Mpc \citep{Tonry:2011aa}. It was classified as a SLSN IIn based on its peak absolute magnitude of -20.7 mag \citep{Prieto:2017aa}. A RHD simulation of the evolution of SN ejecta was performed with HERACLES using a progenitor star similar to that of SN 2017hcc. The profiles at several different checkpoints during the RHD evolution of the SN ejecta were kindly provided by Luc Dessart, along with the synthetic spectra predicted by CMFGEN (private communication). It is an engineered model with 10 \msun \ ejecta mass and 5.7 \msun \ CSM mass with mass loss of 0.2 \msun /year. The radius at the interface between the ejecta and the CSM at the start of the RHD simulation is $10^{15}$ cm. The maximum velocity of  ejecta is 16,600 km/s and the velocity of the CSM is 100 km/s. The RHD calculations with HERACLES do not include the radioctive decay of $^{56}$Ni. As the power from interaction is more dominant compared to radioactive decay, this shall not affect the results. The profiles are truncated at the HERACLES predicted electron scattering optical depth of 30. Here, we present the results of the synthetic spectrum produced with \texttt{SuperLite} by post-processing the checkpoint profile at 36~days from the time of explosion for this model and compare it with the same predicted by CMFGEN. The profile consists for 80 spatial cells. For the \texttt{SuperLite} simulation, 6000 wavelength groups are used in the range of 1 to 30,000~\text{\AA}. The profiles from HERACLES did not include the radiation temperature, so it was set equal to the electron temperature in our simulations.

\begin{figure*}[ht!]
\begin{center}
\gridline{\fig{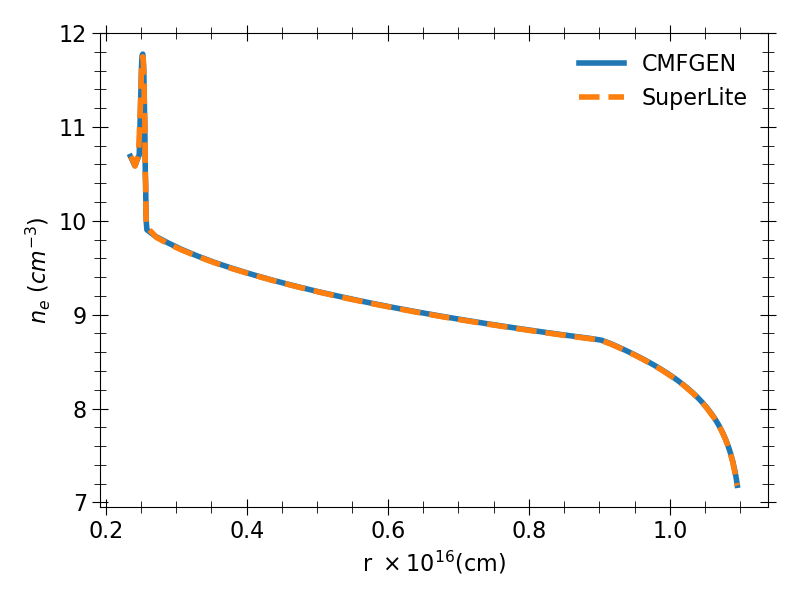}{0.45\textwidth}{(a)}
          	   \fig{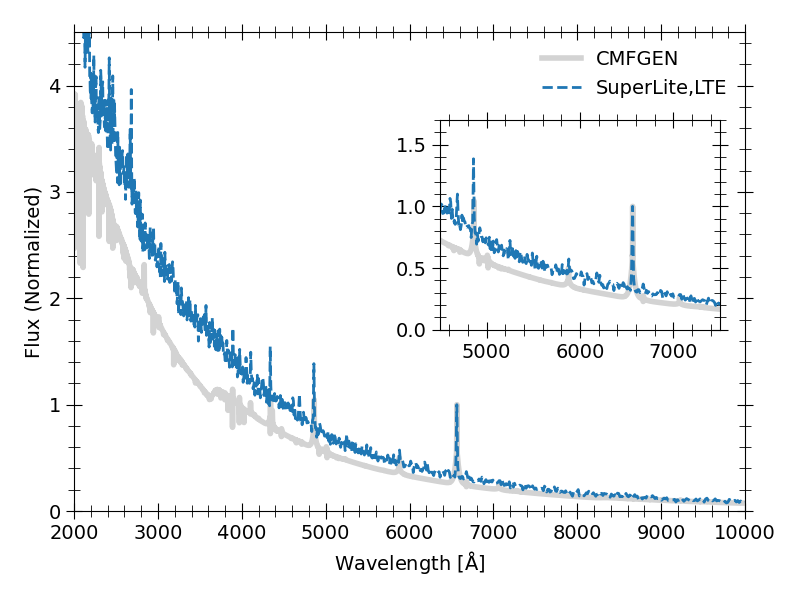}{0.45\textwidth}{(b)}}
\caption{Panel (a) shows a comparison of the electron number density calculated with the \texttt{SuperLite} code under LTE assumption with the CMFGEN code for a SN IIn model provided by Luc Dessart (private communication), both of which show a good agreement. Panel (b) shows a comparison of the spectrum produced with \texttt{SuperLite} under LTE conditions (blue) to that produced by CMFGEN (grey) for the same model. The spectra are normalized to the H$\alpha$ line strength for comparison. The two codes show a good agreement. \label{fig:Dess_36d}}
\end{center}
\end{figure*}

The electron density calculated with \texttt{SuperLite} using LTE assumption qualitatively matches that predicted by CMFGEN as shown in the left panel of Figure~\ref{fig:Dess_36d}.
In the right panel of Figure~\ref{fig:Dess_36d}, the spectrum predicted by \texttt{SuperLite} using LTE assumption at 36 days since explosion is compared with that predicted by CMFGEN. The line ratios in this plot show that the H$\alpha$ line is comparable in strength to the H$\beta$ line in the spectrum predicted by \texttt{SuperLite}, while the H$\alpha$ line is stronger in CMFGEN spectrum. The differences in the line ratios and continuum emission can be attributed to the lack of radiation temperature structure in the input profile, as well as the LTE assumption. We are actively working on resolving these discrepancies. We would like to note, however, that the spectra generated with CMFGEN and \texttt{SuperLite} are otherwise similar.

\section{Results and Conclusions} \label{sec:results}

In this paper we have presented the first publicly-available, Monte Carlo radiation transport code, \texttt{SuperLite}, that is optimized to compute synthetic spectra for astrophysical transients strongly affected by circumstellar interactions (CSI). \texttt{SuperLite} enables users to post-process the hydrodynamic profiles of interacting transient events with non-homologously expanding outflows. \texttt{SuperLite} is published\footnote{https://github.com/gururajw/superlite} under GNU3 public license and it is free to use and develop. The \texttt{SuperLite} code is developed by significantly modifying the existing \texttt{SuperNu} code \citep{Wollaeger:2013aa,Wollaeger:2014aa}. \texttt{SuperLite} uses the multigroup IMC-DDMC approach with accelerated diffusion theory and relativistic corrections of the first order. It has Message-Passing Interface (MPI) and OpenMP capabilities for parallelization of large grids, which allows faster computations. The code uses standard LTE assumptions to calculate the opacities in each domain cell that include the bound-bound opacities for several thousand lines up to $Z = 30$ and analytic fits for bound-free and free-free opacities. The standard elastic Thomson scattering opacity is also included in the calculations. It has the following enhancements that distinguish the code from its parent code and some other publicly-available SN spectral synthesis codes: 
\begin{itemize}
    \item The commonly used assumption of the SN ejecta to be homologously expanding (i.e. $v\propto r$ at any given time $t$ since explosion) is not made in the \texttt{SuperLite} code. The equations such as the transport equation to advance the MC paricles and the Doppler shift corrections are solved in their original form without assuming homology. This is important for the outflows affected by CSI where the assumption of homology breaks.
    \item We have also implemented and are currently testing the non-LTE treatment for calculating the b-b opacities for several hundred hydrogen lines using the rate coefficients that include radiative recombination and photoionization rates, and electron-impact excitation and de-excitation rates for excitation states of singly ionized hydrogen up to the principal quantum number of $n=10$.
\end{itemize}   

We have tested \texttt{SuperLite} by performing some standard MCRT tests, as well as, by successfully simulating spectra for typical SNe of different types as demonstrated in sections \ref{sec:verification} \& \ref{sec:SN_applications}. We have compared the spectra simulated by our code to that simulated by its parent code \texttt{SuperNu} for a SN Ia and SN IIP with homologous outflows. We have also compared the the spectra simulated by our code for a SN IIP to the observations of a similar SN, and for a SN IIn to the calculations performed by the CMFGEN code. Our code compares well with the other codes and observations for these cases. 

With \texttt{SuperLite}, we can model a multitude of transients powered by strong CSI, such as SLSNe, PPISN shell collisions, collisions with material expelled by a past merger and can model luminous, rare and uncharacteristic transient phenomena. This will help with interpretation of transient events discovered by the current and next generation high-cadence, wide-field transient surveys conducted by facilities such as the Vera Rubin Telescope, ZTF and others by enabing direct comparisons to observed time-series of transient spectra. 

\texttt{SuperLite}, being a Monte Carlo code, is easily extendable to work with 2-D and 3-D geometries. We plan to adapt \texttt{SuperLite} to multi-D in the near future, which will enable it to model spectra for transient events interacting with non spherically--symmetric circumstellar environments and help us explore viewing angle effects. Furthermore, a new implementation that implicitly couples a hydrodynamic solver with MCRT will be introduced to properly model the feedback of radiation into the expanding material, similar to that presented by \citet{Noebauer:2019aa,Roth:2015aa}. This will further enhance the capability of \texttt{SuperLite} as an open-source radiation transport tool to accelerate the interpretation of new intriguing transient astrophysical phenomena that challenge our understanding of extreme cosmic catastrophes.

\begin{acknowledgments}
We thank Luc Dessart, Sanjana Curtis for kindly providing us their model inputs to test \texttt{SuperLite} code. We would like to thank J. Craig Wheeler, Chris Fryer, Nathan Smith, David Branch for useful discussions. GW and EC would like to thank the National Science Foundation (NSF) for their support made possible by the NSF grant AST-1907617.
\end{acknowledgments}

%



\software{\texttt{SuperNu} \citep{Wollaeger:2013aa,Wollaeger:2014aa},
          STELLA \citep{Blinnikov:1993aa,Blinnikov:1998aa,Blinnikov:2004aa,Blinnikov:2006aa},
          MESA \citep{Paxton:2011aa,Paxton:2013aa,Paxton:2015aa,Paxton:2018aa,Paxton:2019aa},
          }









\bibliography{superlite.bib}{}
\bibliographystyle{aasjournal}



\end{document}